\documentclass[10pt,twocolumn]{IEEEtran}

\usepackage{graphicx}
\usepackage{amsmath}
\usepackage{amssymb}
\usepackage[caption=false]{subfig}
\usepackage[noadjust]{cite}
\usepackage{float}
\usepackage{algorithm}
\usepackage{bibentry}
\usepackage{balance}
\usepackage{algorithm}
\usepackage{algorithmicx}
\usepackage{algpseudocode}
\usepackage{xcolor}
\usepackage{subfig}
\usepackage{wrapfig}

\graphicspath{{img/}}

\begin{document}

\bstctlcite{IEEEexample:BSTcontrol}


\title{Precoder Design for Physical-Layer Security and Authentication in Massive MIMO UAV Communications
\thanks{This work is supported in part by the INL Laboratory Directed Research Development (LDRD) Program under DOE Idaho Operations Office Contract DEAC07-05ID14517. An earlier version of this work has appeared in~\cite{maeng2020precoder}.}\thanks{S. J. Maeng, \.{I}. G\"{u}ven\c{c}, and H. Dai are with the Department of Electrical and Computer Engineering, North Carolina State University, Raleigh, NC 27606 USA (e-mail: smaeng@ncsu.edu; madeshmu@ncsu.edu; iguvenc@ncsu.edu; hdai@ncsu.edu).}\thanks{Yavuz Yap{\i}c{\i} is with Qualcomm Inc., San Diego, CA 92121 USA (e-mail: yyapici@qti.qualcomm.com).}\thanks{A. Bhuyan is with the INL Wireless Security Institue, Idaho National Laboratory, Idaho Falls, ID 83402 USA (e-mail: arupjyoti.bhuyan@inl.gov).}}

\author{\IEEEauthorblockN{Sung Joon Maeng, Yavuz Yap{\i}c{\i}, \textit{Senior Member, IEEE}, \.{I}smail G\"{u}ven\c{c}, \textit{Fellow, IEEE}, Arupjyoti Bhuyan, \textit{Senior Member, IEEE} and Huaiyu Dai, \textit{Fellow, IEEE}}}%

\maketitle

\begin{abstract}
Supporting reliable and seamless wireless connectivity for unmanned aerial vehicles (UAVs) has recently become a critical requirement to enable various different use cases of UAVs. Due to their widespread deployment footprint, cellular networks can support beyond visual line of sight (BVLOS) communications for UAVs.  In this paper, we consider cellular connected UAVs (C-UAVs) that are served by massive multiple-input-multiple-output (MIMO) links to extend coverage range, while also improving  physical layer security and authentication. We consider Rician channel and propose a novel linear precoder design for transmitting data and artificial noise (AN). We derive the closed-form expression of the ergodic secrecy rate of C-UAVs for both conventional and proposed precoder designs. In addition, we obtain the optimal power splitting factor that divides the power between data and AN by asymptotic analysis. Then, we apply the proposed precoder design in the fingerprint embedding authentication framework, where the goal is to minimize the probability of detection of the authentication tag at an eavesdropper. In simulation results, we show the superiority of the proposed precoder in both secrecy rate and the authentication probability considering moderate and  large number of antenna massive MIMO scenarios.
\end{abstract}


\section{Introduction}

In recent years, unmanned aerial vehicles (UAVs) received considerable attention as a promising future technology for  various different use cases. Applications of UAVs include  monitoring and surveillance for military missions, search and rescue, package  delivery, and broadcasting of live video for commercial uses \cite{marojevic2020advanced}. Furthermore, a UAV can be deployed in the sky as a flying mobile base station (BS) for improving reliability and flexibility in  cellular networks~\cite{rupasinghe2018non}. To support such diverse potential applications, high-throughput, low-latency, and long-range connectivity are essential. Massive multiple-input-multiple-output (MIMO) communications is one of the key technologies that can  support high and stable throughput by using large number of antennas on the BS~\cite{chandhar2017massive}. In this paper, we consider cellular-connected UAVs (C-UAVs) that are served by the massive MIMO technology from BSs on the ground for improved coverage. 

Security is a highly critical aspect of wireless communications. Traditionally, security in communications is established by cryptographic encryption techniques at the application layer, which relies on computing power limitation for the decryption. Physical-layer security concepts have received more attention since the wire-tap channel is introduced in \cite{wyner1975wire}. In that work, information-theoretic secrecy, known as a secrecy capacity, is defined by the maximum rate that the legitimate user can achieve while the eavesdropper is not able to decode the message. Secret communication by generating artificial noise (AN) in the MIMO system is first studied in \cite{negi2005secret}, and is followed up with several follow up works over the past decades such as \cite{goel2008guaranteeing}, \cite{nguyen2018secure}.
The achievable secrecy rate was evaluated in \cite{zhu2014secure, zhu2015linear} considering various precoder design schemes.
Data precoder is designed by the multi-user linear precoding such as matched-filter (MF), zero-forcing (ZF), and regularized channel inversion (RCI), while AN is precoded by either null-space precoding or Gaussian random vector generation. The allocated power between message and AN is also optimized by the maximum secrecy rate. In \cite{wang2015jamming},  directional jamming in the Rician channel is proposed and its performance is compared with that of the uniform jamming. In \cite{geraci2012secrecy}, the RCI precoding is optimized for the secrecy in massive MIMO systems without AN transmission. Hybrid structure precoding, which splits analog and digital parts in precoder design for secure transmission is studied in order to reduce the hardware complexity with secrecy performance loss \cite{zhu2016secure}.

The existing literature on secure communications for UAV networks is mostly focused on the UAV trajectory design and power control. In \cite{cui2018robust, zhang2019securing}, the trajectory and the transmit power are jointly optimized by the secrecy rate. A jamming UAV is considered in \cite{zhou2019uav}, and user scheduling is jointly optimized in \cite{lee2018uav}. However, the precoder design for the UAV massive MIMO system is rarely studied. In the above papers, the location information of the passive eavesdropper, whether it is perfect or imperfect, is utilized in designing the trajectory and optimizing the transmit power. In this sense, we adopt the similar assumption of the passive UAV eavesdropper (UAV-Eve) and assume that the ground station (GS) is able to obtain the imperfect location information of the UAV-Eve in designing precoders.

In addition to maintaining secure communications with UAVs, accurate authentication of the UAVs carries critical importance to establish the communication link in the first place, and we will tackle this problem jointly with secure communications. Fingerprint embedding authentication framework is a physical layer authentication that distinguishes the identity of a message while denying the impersonation attacks from the eavesdropper \cite{paul2015wireless}. In this framework, the transmitter superimposes the low-power authentication tag on the data, and the tag is encrypted by the secret key. The intended receiver authenticates the tag by using an already shared key, while the attacker tries to guess the correct key by the received signal. The probability that the attacker successfully guesses the secret key is a typical performance metric to characterizes the vulnerability of the authentication framework. 
To our best knowledge, the fingerprinting authentication is introduced and applied in a MIMO system in \cite{paul2011mimo}, while in \cite{verma2015physical}, the fingerprint embedding framework is validated by single-antenna software defined radio (SDR) experiments. In \cite{perazzone2019fingerprint}, the AN is introduced in the authentication framework in the multiple-input single-output (MISO) system. In \cite{perazzone2021artificial}, the imperfect channel state information (CSI) is considered with AN in a MIMO system. However, to our best knowledge, fingerprinting authentication on a multi-user MIMO system as well as with UAVs have not been studied yet.

In our previous work \cite{maeng2020precoder}, we focus on designing various linear and non-linear precoders for a millimeter wave UAV-BS serving ground users while also minimizing information leakage to eavesdroppers. Our new work presented in this paper focuses on the linear precoder design while deriving closed-form expression of the ergodic secrecy rate, considering the Rician channel model and large number of antennas regime. We now consider a  ground BS serving to UAV users, and we analytically show the superiority of our proposed precoder design over existing techniques. The contributions of our work can be listed as follows.
\begin{itemize}
  \item[--] We propose the data and AN precoder based on ZF precoding with the imperfect location information of UAV-Eve. We compare it with the conventional approach where the data precoder is designed by ZF precoding and AN precoder is designed by null-space precoding \cite{zhu2015linear}.
  \item[--] We model calibration error of the elevation angle of the line-of-sight (LoS) by the real-valued Gaussian random variable and derive the mean square error (MSE) of LoS component channel by an approximation to show the dependency of the parameters.
  \item[--] We consider the Rician channel model and express the closed-form expression of the ergodic secrecy rate for both the conventional and the proposed precoder designs. We also show the large antennas and high Rician K-factor limit on the ergodic secrecy rate. Many analytical derivations refers to
  \cite{zhang2014power,zhu2014secure}.
  \item[--] We find the optimal power splitting factor that maximizes the ergodic secrecy rate by the large antennas analysis. 
  \item[--] We adopt the fingerprint embedding authentication framework and optimize the tag power factor. We show that the proposed precoder design outperforms the conventional precoder design in the authentication framework. 
  \end{itemize}
  
\begin{figure}[t]
	\centering
	\vspace{-0.0in}
	\includegraphics[width=0.5\textwidth]{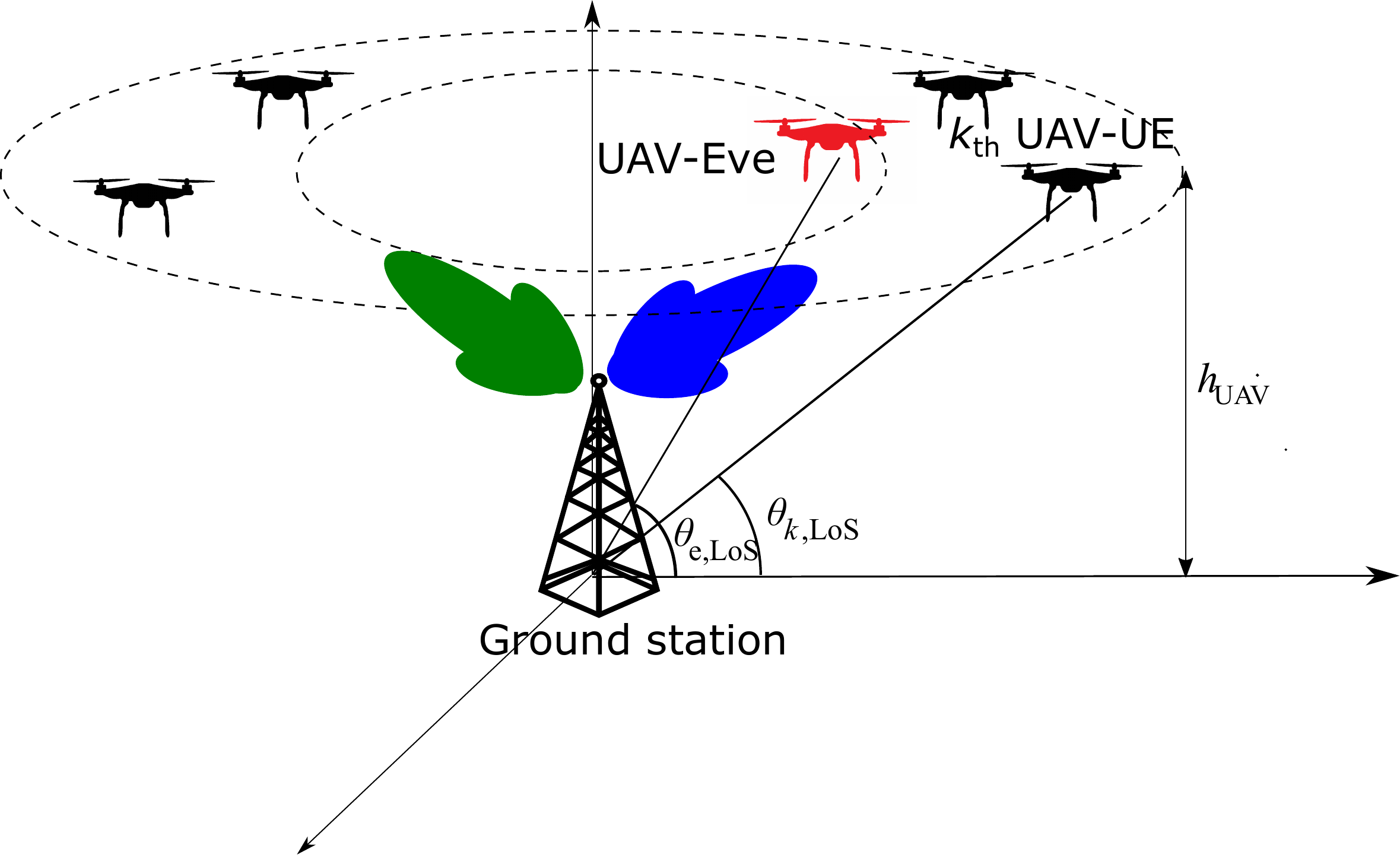}
	\caption{Illustration of the system model with a ground station, multiple UAV-UEs and single UAV-Eve.} 
	\label{fig:illu}
\end{figure}
  
\section{System Model} \label{sec:system}

In this section, we present the system design of the UAV cellular networks as shown in Fig.~\ref{fig:illu}. We consider a single GS with multiple UAV users (UAV-UEs) and single UAV-Eve in scenario, and the multiple-input single-output single-antenna-eavesdropper (MISOSE) type of wiretap channels \cite{lin2013secrecy}. The GS equipped with $N_{\rm t}$ antennas transmits zero-mean unit variance complex data symbols ($\textbf{s}_k$) toward $K$ single antenna equipped UAV-UEs through the designed multi-user MIMO (MU-MIMO) precoder (\textbf{W}). On the other hand, a UAV-Eve attempts to monitor the data and illegally pretends to be the GS. To protect the data, the GS transmits zero-mean unit variance complex Gaussian distributed AN ($\textbf{z}_i$) with $N_{\mathsf{AN}}$ dimensions AN precoder design ($\textbf{V}$) as well. Furthermore, the low power authentication tag ($\textbf{t}_k$) is superimposed on the data and simultaneously transmitted for the purpose of security. 

Then, we can formulate the received signal of the UAV-UEs as follows:
\begin{align}\label{eq:received_signal:user}
        \textbf{y}_u^{\rm H}&=\sum_{k=1}^{K}\frac{\sqrt{\mathsf{P}_\mathsf{Tx}}}{\sqrt{\mathsf{PL}_u}}\textbf{h}_u^{\rm H}\sqrt{\phi}\textbf{w}_k(\sqrt{(1-\delta)}\textbf{s}_k+\sqrt{\delta}\textbf{t}_k)^{\rm H}\nonumber\\
        &+\sum_{i=1}^{N_{\mathsf{AN}}}\frac{\sqrt{\mathsf{P}_\mathsf{Tx}}}{\sqrt{\mathsf{PL}_u}}\textbf{h}_u^{\rm H}\sqrt{1-\phi}\textbf{v}_i\textbf{z}_i^{\rm H}+\textbf{n}_u^{\rm H},
\end{align}
where $\textbf{y}_u$ indicates the received signal of the  $u_{\rm th}$ UAV-UE, $\mathsf{P}_\mathsf{Tx}$ is transmit power, $\mathsf{PL}_u$ denotes path-loss of the $u_{\rm th}$ UAV-UE, $\textbf{h}_u$ is small-scale fading channel of the $u_{\rm th}$ UAV-UE, $\textbf{z}_i$ is zero-mean unit variance complex Gaussian AN symbols, and $\textbf{n}_u$ is additive complex Gaussian noise of the $u_{\rm th}$ UAV-UE whose entries follow $\mathcal{CN}(0,\sigma_n^2)$. In addition, the power of precoders $\textbf{W}$, $\textbf{V}$ are spilt by the power splitting factor ($0\leq\phi\leq1$), and the power of $\textbf{t}_k$ is allocated by the tag power factor ($0\leq\delta\leq1$).

\begin{figure}[t]
	\centering
	\vspace{-0.0in}
	\includegraphics[width=0.49\textwidth]{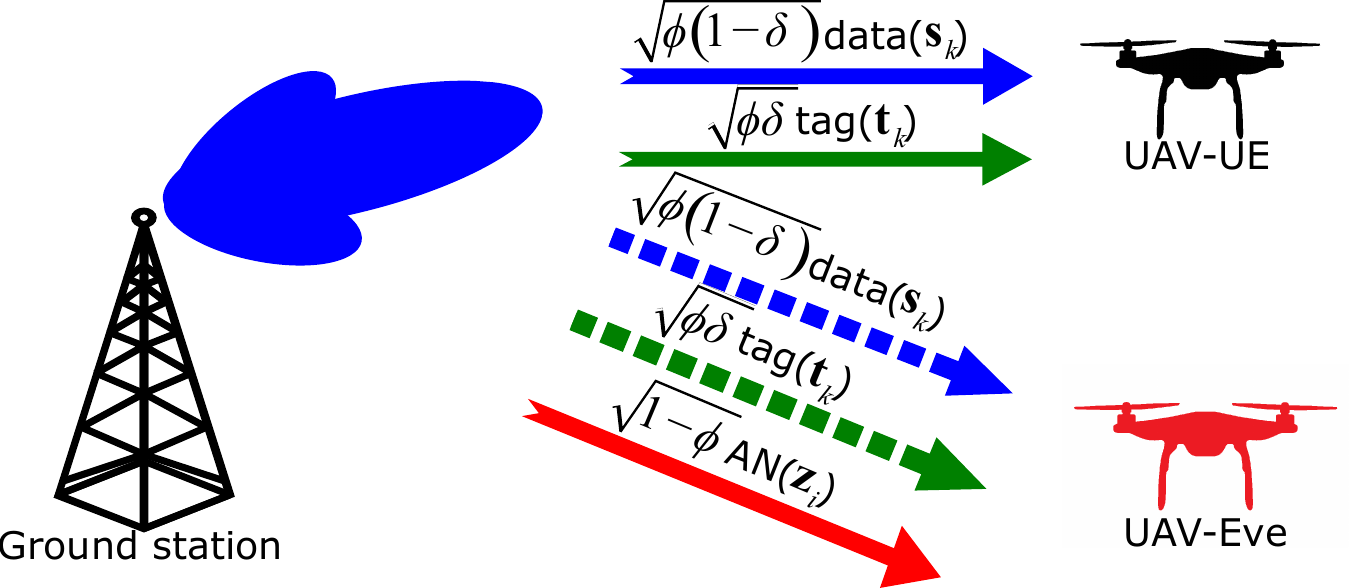}
	\caption{Illustration of flows of different streams of signals to the UAV-UE and the UAV-Eve. The power splitting factor (for AN) and the tag power factor (for authentication tag) are denoted by $\phi$ and $\delta$, respectively.} 
	\label{fig:illu2}
\end{figure}

Similarly, the received signal of the UAV-Eve is given by
\begin{align}\label{eq:received_signal:Eve}
        \textbf{y}_{\mathsf{e}}^{\rm H}&=\sum_{k=1}^{K}\frac{\sqrt{\mathsf{P}_\mathsf{Tx}}}{\sqrt{\mathsf{PL}_{\mathsf{e}}}}\textbf{h}_{\mathsf{e}}^{\rm H}\sqrt{\phi}\textbf{w}_k(\sqrt{(1-\delta)}\textbf{s}_k+\sqrt{\delta}\textbf{t}_k)^{\rm H}\nonumber\\
        &+\sum_{i=1}^{N_{\mathsf{AN}}}\frac{\sqrt{\mathsf{P}_\mathsf{Tx}}}{\sqrt{\mathsf{PL}_{\mathsf{e}}}}\textbf{h}_{\mathsf{e}}^{\rm H}\sqrt{1-\phi}\textbf{v}_i\textbf{z}_i^{\rm H}+\textbf{n}_{\mathsf{e}}^{\rm H},
\end{align}
where the subscript of `$\mathsf{e}
$' indicates the component of UAV-Eve; e.g. $\textbf{h}_{\mathsf{e}}$ denotes the small-scale fading channel that corresponds to the UAV-Eve. In Fig.~\ref{fig:illu2}, we show the flows of different steams of signals to a UAV-UE and the UAV-Eve respectively.

\subsection{Channel Model and the Location of UAVs in the Network}

We adopt Rician fading channel for the ground-to-air channel between the GS and UAV-UEs, and between the GS and the UAV-Eve. It is well-known that the LoS is easily secured in ground-to-air propagation due to the height of the aerial objects \cite{ khawaja2019survey, zeng2019accessing}. The deterministic LoS component ($\textbf{h}_{\mathsf{LoS},k}$) and the scattered random component ($\textbf{h}_{\mathsf{NLoS},k}$) are mixed by Rician K-factor ($\kappa$) as \cite{wang2015jamming}
\begin{align}\label{eq:Rician_user}
    \textbf{h}_k = \sqrt{ \frac{ \kappa }{ \kappa + 1 } }\textbf{h}_{\mathsf{LoS},k}+\sqrt{ \frac{ 1 }{ \kappa + 1 }}\textbf{h}_{\mathsf{NLoS},k},
\end{align}
where the entries of  $\textbf{h}_{\mathsf{NLoS},k}$ is independent and identically distributed (i.i.d.) complex Gaussian random variable $\sim\mathcal{CN}(0,1)$. The LOS component,  $\textbf{h}_{\mathsf{LoS},k}$ can be represented by the steering vector as follows:
\begin{align}\label{eq:h_LoS:user}
    \textbf{h}_{\mathsf{LoS},k}&=\textbf{a}_{N_{\mathsf{t}}}(\theta_{k,\mathsf{LoS}}),
\end{align}
where $\theta_{k,\mathsf{LoS}}$ indicates the elevation angle of the LoS of the $k_{\rm th}$ UAV-UE as shown in Fig.~\ref{fig:illu}. We consider that the vertical oriented uniform linear array (ULA) antenna and the steering vector are given by
\begin{align}\label{eq:steering_vector}
    \textbf{a}_{N}(\theta)&=\left[1\; e^{-j\frac{2\pi d_{\rm s}}{\lambda}\sin\theta}\;\dots\; e^{-j\frac{2\pi d_{\rm s}}{\lambda}(N-1)\sin\theta}\right]^{\rm T},
\end{align}
where $d_{\rm s}$, $\lambda$ denote antenna spacing and wave-length. The dB scale path-loss is calculated by 3GPP urban micro (UMi) environment given as \cite{3gpp}
\begin{align}\label{eq:PL}
    \mathsf{PL}_k  &= 32.4 + 21\log_{10}\left( d_{k,\mathsf{LoS}}\right) + 20\log_{10}\left( f_{\rm c} \right),
\end{align}
where $d_{k,\mathsf{LoS}}$ is the LoS distance of the $k_{\rm th}$ UAV-UE, which can be calculated by the horizontal distance ($d_k$) and the UAV height ($h_{\mathsf{UAV}}$) as $d_{k,\mathsf{LoS}}=\sqrt{d_k^2 + h_{\mathsf{UAV}}^2 }$, and $f_{\rm c}$ denotes the normalized carrier frequency by 1 GHz. By the same way, we can express the small-scale fading channel of the UAV-Eve as \begin{align}\label{eq:Rician_eve}
    \textbf{h}_{\mathsf{e}} = \sqrt{ \frac{ \kappa }{ \kappa + 1 } }\textbf{h}_{\mathsf{LoS},\mathsf{e}}+\sqrt{ \frac{ 1 }{ \kappa + 1 }}\textbf{h}_{\mathsf{NLoS},\mathsf{e}}.
\end{align}

We assume that all UAVs' height are fixed by the same altitude and the horizontal distance of UAV-UE ($d_k$) follows uniform distribution $\sim\mathcal{U}[d_{\mathsf{min}},d_{\mathsf{max}}]$. In addition, the horizontal distance of the UAV-Eve ($d_{\mathsf{e}}$) is fixed at $d_{\mathsf{min}}$, which is the securely 
the vulnerable case that the distance of the UAV-Eve is closer to the GS than UAV-UEs.

\subsection{Channel Knowledge Assumption of Ground Station}

In this subsection, we discuss the channel assumption in designing the precoders by the GS. We assume that the CSI of the UAV-UEs is perfectly known by the GS, while the imperfect elevation angle information of the UAV-Eve is known by the GS. We assume that the GS is capable of detecting the location of the UAV-Eve; this may be possible through techniques such as RF sensing, cameras, or radar, as was recently studied in the literature~\cite{shang2019unmanned,guvenc2018detection,ezuma2019detection}. The GS may also detect the UAV-Eve by monitoring the power leakage from the eavesdropper~\cite{mukherjee2012detecting}. The LoS elevation angle of the UAV-Eve can be expressed as
\begin{align}\label{eq:LoS_angle}
    \theta_{\rm e,\mathsf{LoS}}&=\arctan(\frac{h_{\mathsf{UAV}}}{d_{\rm e}}).
\end{align}
Then, the imperfect LoS elevation angle of the UAV-Eve can be written as
\begin{align}\label{eq:LoS_angle}
    \hat{\theta}_{\rm e,\mathsf{LoS}}&=\theta_{\rm e,\mathsf{LoS}}+\epsilon,
\end{align}
where $\epsilon$ indicates the angular calibration error (the unit is degree) following zero-mean  Gaussian distributed real-valued random variable $\mathcal{N}(0,\sigma^2_{\epsilon})$.

\subsection{Secrecy Rate for UAV-UE}
We can express signal-to-interference-plus-noise ratio (SINR) of the UAV-UE using \eqref{eq:received_signal:user} as follows
\begin{align}\label{eq:SINR_u}
    \!\!\mathsf{SINR}_u = \frac{\phi(1-\delta)\textbf{w}_u^{\rm H}\textbf{h}_u\textbf{h}_u^{\rm H}\textbf{w}_u}{\sum_{k\neq u}^{K} \phi\textbf{w}_k^{\rm H}\textbf{h}_u\textbf{h}_u^{\rm H}\textbf{w}_k + \sum_{i=1}^{N_{\mathsf{AN}}} (1-\phi)\textbf{v}_i^{\rm H}\textbf{h}_u\textbf{h}_u^{\rm H}\textbf{v}_i + \rho^{{-}1}_u},
\end{align}
where $\rho_u=\frac{\mathsf{P}_\mathsf{Tx}}{\mathsf{PL}_u\sigma_{n}^2}$. Similarly, the SINR of the UAV-Eve can be written by \eqref{eq:received_signal:Eve} as follow
\begin{align}\label{eq:SINR_eve}
    \!\!\mathsf{SINR}_\mathsf{e} = \frac{\phi(1-\delta)\textbf{w}_u^{\rm H}\textbf{h}_\mathsf{e}\textbf{h}_\mathsf{e}^{\rm H}\textbf{w}_u}{\sum_{i=1}^{N_{\mathsf{AN}}} (1-\phi)\textbf{v}_i^{\rm H}\textbf{h}_\mathsf{e}\textbf{h}_\mathsf{e}^{\rm H}\textbf{v}_i + \rho^{{-}1}_\mathsf{e}},
\end{align}
where $\rho_{\mathsf{e}}=\frac{\mathsf{P}_\mathsf{Tx}}{\mathsf{PL}_{\mathsf{e}}\sigma_{n}^2}$. We consider the worst case assumption for the SINR formulation of the UAV-Eve that the UAV-Eve is able to fully decode and eliminate contributions from the other UAV-UEs \cite{zhu2014secure}. Then, the ergodic secrecy rate can be written by the above SINR expressions as \cite{zhu2014secure}
\begin{align}\label{eq:secrecy_rate}
     \mathsf{R}^\mathsf{sec}_u &= \left[\mathbb{E}\left[\log_2(1+\mathsf{SINR}_u)\right]-\mathbb{E}\left[\log_2(1+\mathsf{SINR}_\mathsf{e})\right]\right]^+\\
     &=\left[\mathsf{R}_u-\mathsf{R}_{\mathsf{e}}\right]^+,
\end{align}
where $\left[x\right]^+=\max(0,x)$.

\section{Precoder Design}

In this section, we design the MU-MIMO precoder ($\textbf{W}$) for the data, as well as the precoder for the AN ($\textbf{V}$). We first introduce already established conventional precoder design that utilizes only UAV-UEs CSI. Then, we propose a precoder design that takes into account the limited CSI of UAV-Eve as well as the CSIs of the UAV-UEs. Our proposed precoder design uses the location information of the UAV-Eve, which brings the degree of freedom to improve the secrecy performance.

\subsection{Conventional Precoder Design without UAV-Eve CSI}

The secure precoder design for the massive MIMO system is initially studied in \cite{zhu2014secure}. The authors design the MF precoder for the data, and the random matrix and the null-space precoding are designed for the AN precoder. In another work, \cite{zhu2015linear} extends the study of the linear data precoders to the ZF and the RCI precoders. In this paper, we consider the ZF precoder for the MU-MIMO data precoder and the null-space precoding for AN precoder. This is the smart way to design the precoders without the eavesdropper information for secure transmission. The ZF precoder eliminates the interference from the other users' data, while the AN precoder transmits noise that is orthogonal to the users' channel such that it is canceled out at the users. In this manner, the AN does not degrade UAV-UEs' channel, while it degrades the UAV-Eve's channel.

The ZF design for the MU-MIMO precoder is given by
\begin{align}\label{eq:ZF_precoder:conv}
    \Tilde{\textbf{W}} &= \textbf{H}(\textbf{H}^{\rm H}\textbf{H})^{{-}1},
\end{align}
where $\textbf{H}=[\textbf{h}_1,\dots,\textbf{h}_K]$ is the aggregate channel matrix of $K$ UAV-UEs. The power of each column vector of precoder $\Tilde{\textbf{W}}=[\Tilde{\textbf{w}}_1,\dots,\Tilde{\textbf{w}}_K]$ is uniformly allocated by making it satisfy the power constraint $\|\textbf{W} \|^2 = 1$ as
\begin{align} \label{eq:ZF_precoder_norm:conv}
    \textbf{w}_k&=\frac{1}{\sqrt{K}\|\Tilde{\textbf{w}}_k\|}\Tilde{\textbf{w}}_k.
\end{align}
The null-space AN precoder can be expressed as
\begin{align} \label{eq:AN_precoder:conv}
    \Tilde{\textbf{V}}&=\text{null}(\textbf{H}^{\rm H}),
\end{align}
where each column vector of $\Tilde{\textbf{V}}=[\Tilde{\textbf{v}}_1,\dots,\Tilde{\textbf{v}}_{N_{\mathsf{AN}}}]$ is orthogonal to the users channel matrix, and the maximum number of $N_{\mathsf{AN}}$ satisfies $N_{\mathsf{AN}}=N_{\mathsf{t}}-K$. Similarly, the uniform power allocation of precoder is applied with the power constraint $\|\textbf{V} \|^2 = 1$ as 
\begin{align} \label{eq:AN_precoder_norm:conv}
    \textbf{v}_i &= \frac{1}{\sqrt{N_{\mathsf{AN}}}\|\Tilde{\textbf{v}}_i\|}\Tilde{\textbf{v}}_i.
\end{align}

The above MU-MIMO precoder and AN precoder design are used as a reference in order to compare it with the proposed precoder design that is described in the next subsection.

\subsection{Proposed Precoder Design with Limited UAV-Eve CSI}

We propose the MU-MIMO data precoder and AN precoder with the elevation angle information of the UAV-Eve ($\hat{\theta}_{\rm e,\mathsf{LoS}}$) that the GS obtains. The location information of the eavesdropper with the potential error is already used in the UAV trajectory design for secure communications in several papers \cite{cui2018robust,zhang2019securing}. Unlike the conventional AN precoder design that broadcasts the jamming noise to the multiple dimensional spaces, we design the precoder such that the energy is focused on the direction of the UAV-Eve. Furthermore, the data precoder is designed such that it is null to the direction of the UAV-Eve. By doing so, we can considerably suppress the quality of the signal on the UAV-Eve. The benefit of directional jamming compared with uniform jamming is studied in \cite{wang2015jamming}. However, the paper designs the directional jamming based on the null space precoding and selecting a few good column vectors out of the total, but not based on the actual direction of the eavesdropper. 

Our proposed precoder designs is described as follows. We can express the directional vector of the UAV-Eve as
\begin{align}\label{eq:directional_Eve}
    \textbf{g}_{\rm e}&=\textbf{a}_{N_{\mathsf{t}}}(\hat{\theta}_{\rm e,\mathsf{LoS}}).
\end{align}
Treating $\textbf{g}_{\rm e}$ as if it is a regular user, the ZF precoder is given as follows: 
\begin{align}\label{eq:ZF_precoder:prop}
    \textbf{F}&=\textbf{G}(\textbf{G}^{\rm H}\textbf{G})^{-1},
\end{align}
where $\textbf{G}=[\textbf{H}\; \textbf{g}_{\rm e}]$ is a virtual channel matrix which aggregates the UAV-UEs channel and the UAV-Eve's directional vector. By this manner, the columns vectors $\textbf{f}_1, \dots, \textbf{f}_K$ nullify not only other users but also the direction of UAV-Eve. Besides, the columns vector $\textbf{f}_{K+1}$ is orthogonal to $K$ UAV-UEs while it correlates with the direction of the UAV-Eve. 

Then, we design the proposed MU-MIMO precoder for UAV-UEs as follows: 
\begin{align}\label{eq:data_precoder:prop}
    \Tilde{\textbf{W}} &= \left[\textbf{f}_1,\dots,\textbf{f}_K\right].
\end{align}
We apply the uniform power allocation by $\eqref{eq:ZF_precoder_norm:conv}$ to obtain $\textbf{W}$. Moreover, the proposed AN precoder is given by
\begin{align}\label{eq:AN_precoder:prop}
    \Tilde{\textbf{v}} &= \textbf{f}_{K+1}.
\end{align}
Similarly, $\textbf{v}$ is obtained by the uniform power allocation using $\eqref{eq:AN_precoder_norm:conv}$. Note that $N_{\mathsf{AN}}=1$ in the proposed design since it is the single direction.

\section{Analysis of Achievable Secrecy Rate}

In this section, we derive the closed-form equation of the ergodic secrecy rate in \eqref{eq:secrecy_rate} for both the conventional and the proposed precoders designs. We compare the performance of the two precoder designs by the obtained results. In addition, we discuss the effect of the angular calibration error ($\epsilon$) on the ergodic secrecy rate. 

\subsection{Ergodic Achievable Rate of the UAV-UE}

The ergodic achievable rate of the UAV-UE is expressed from \eqref{eq:SINR_u}, \eqref{eq:secrecy_rate} as
\begin{align}\label{eq:rate_user}
    \mathsf{R}_u&=\mathbb{E}\left\{\log_{2}\left(1+\mathsf{SINR}_u\right)\right\}\nonumber\\
    &=\mathbb{E}\left\{\log_{2}\left(1+\frac{\phi(1-\delta)|\textbf{h}_u^{\rm H}\textbf{w}_u|^2}{ \rho^{{-}1}_u}\right)\right\}\nonumber\\
    &\overset{(a)}{\approx}\log_{2}\left(1+\frac{\phi(1-\delta)\mathbb{E}\left\{|\textbf{h}_u^{\rm H}\textbf{w}_u|^2\right\}}{ \rho^{{-}1}_u}\right),
\end{align}
where the approximation $(a)$ comes from the \cite[Lemma 1]{zhang2014power}. Since interference from other users and AN is cancelled out by precoder designs in both the conventional and the proposed designs, the first and the second terms of denominator in $\mathsf{SINR}_u$ in \eqref{eq:SINR_u} is removed in \eqref{eq:rate_user}. 
We denote the conventional MU-MIMO precoder and AN precoder as $\textbf{W}^{\mathsf{cv}}$, $\textbf{V}^{\mathsf{cv}}$, and the proposed precoders as $\textbf{W}^{\mathsf{pp}}$, $\textbf{V}^{\mathsf{pp}}$ for convenience.
\subsubsection{Conventional precoder design}
The closed-form expression of the ergodic achievable of UAV-UE for the conventional precoder design is derived in the following theorem. 

\newtheorem{theorem}{Theorem}
\begin{theorem}\label{t1}
The ergodic achievable rate of UAV-UE in \eqref{eq:rate_user} for the conventional precoder design can be written as
\begin{align}\label{eq:theo1:def}
    \mathsf{R}_u^{\mathsf{cv}}&=\log_{2}\left(1+\frac{(N_{\mathsf{t}}-K)\phi(1-\delta)}{K[\boldsymbol{\Sigma}^{-1}_{\textbf{H}}]_{uu}\rho^{{-}1}_u}\right)~.
\end{align}
\end{theorem}
\begin{IEEEproof}
See Appendix A.
\end{IEEEproof}
\subsubsection{Proposed precoder design}
The closed-form expression of the ergodic achievable of UAV-UE for the proposed precoder design is derived in the following theorem. 
\begin{theorem}
The ergodic achievable rate of UAV-UE for the proposed precoder design can be expressed as
\begin{align}\label{eq:theo2:def}
    \mathsf{R}_u^{\mathsf{pp}}&=\log_{2}\left(1+\frac{(N_{\mathsf{t}}-K-1)\phi(1-\delta)}{K\left([\boldsymbol{\Sigma}^{-1}_{\textbf{H}}]_{uu}+\Delta\right)\rho^{{-}1}_u}\right).
\end{align}
\end{theorem}
\begin{IEEEproof}
See Appendix B.
\end{IEEEproof}
\newtheorem{corollary}{Corollary}
\begin{corollary}
From the closed-form expression of \textit{Theorem 1} and \textit{Theorem 2}, we can conclude that the ergodic achievable rate of UAV-UE for the conventional precoder design is always greater than the proposed precoder design as follow
\begin{align}\label{eq:cor1:def}
    \mathsf{R}_u^{\mathsf{cv}}> \mathsf{R}_u^{\mathsf{pp}}.
\end{align}
\end{corollary}
\begin{IEEEproof}
Comparing \eqref{eq:theo1:def}, \eqref{eq:theo2:def}, we can observe only two different terms between the equations and we can compare them as
\begin{align}
    N_{\mathsf{t}}-K>N_{\mathsf{t}}-K-1,\label{eq:col1:p1}\\
    [\boldsymbol{\Sigma}^{-1}_{\textbf{H}}]_{uu}\leq [\boldsymbol{\Sigma}^{-1}_{\textbf{H}}]_{uu}+\Delta.\label{eq:col1:p2}
\end{align}
From above observations, we can easily obtain \eqref{eq:cor1:def}.
\end{IEEEproof}

Note that the performance degradation of the proposed precoder design results from adding a UAV-Eve directional vector in the ZF precoding such that the precoder needs to make an additional dimension to be orthogonal. However, adding on a single dimension in the ZF precoding does not have much effect, especially when $N_{\mathsf{t}}$ is sufficiently large. Thus, the performance loss in the proposed precoder design is minimal in the large $N_{\mathsf{t}}$. Furthermore, two ergodic achievable rates in \textit{Corollary 1} asymptotically become the same as $N_{\mathsf{t}}$ goes infinity by the following \textit{Corollary 2}.
\begin{corollary}
The ergodic achievable rate of the UAV-UE for the proposed precoder design converges to the conventional precoder design as
\begin{align}\label{eq:cor2:def}
    \mathsf{R}_u^{\mathsf{cv}} - \mathsf{R}_u^{\mathsf{pp}}\to 0,\;\text{as}\;N_{\mathsf{t}}\to\infty.
\end{align}
\end{corollary}
\begin{IEEEproof}
We prove it by showing that expressions on the two sides of the inequalities in \eqref{eq:col1:p1}, \eqref{eq:col1:p2} converge to the same value as $N_{\mathsf{t}}$ grows without bound. Firstly, we can easily get
\begin{align}
    N_{\mathsf{t}}-K\to N_{\mathsf{t}}-K-1,\;\text{as}\;N_{\mathsf{t}}\to\infty.
\end{align}
From \eqref{eq:theo1:p4} and \cite[Corollary 5]{zhang2014power}, we can also get
\begin{align}\label{eq:col2:p2}
    \boldsymbol{\Sigma}_{\textbf{H}}^{-1}\to \textbf{I},\;\text{as}\;N_{\mathsf{t}}\to\infty.
\end{align}
By using \cite[Corollary 5]{zhang2014power}, we can have
\begin{align}\label{eq:col2:p3}
    \frac{1}{N_{\mathsf{t}}}\textbf{H}_{\mathsf{LoS}}^{\rm H}\textbf{g}_{\rm e}\to \textbf{0},\;\text{as}\;N_{\mathsf{t}}\to\infty.
\end{align}
From \eqref{eq:col2:p3}, we can also get
\begin{align}\label{eq:col2:p4}
    \textbf{b}\to \textbf{0},\;\textbf{c}\to \textbf{0},\;\text{as}\;N_{\mathsf{t}}\to\infty.
\end{align}
Then, we can plug \eqref{eq:col2:p2}, \eqref{eq:col2:p4} in \eqref{eq:theo2:p5} to show
\begin{align}\label{eq:col2:p5}
    \Delta\to 0,\;\text{as}\;N_{\mathsf{t}}\to\infty.
\end{align}
Finally, using \eqref{eq:col2:p5} we can obtain
\begin{align}\label{eq:col2:p6}
    [\boldsymbol{\Sigma}^{-1}_{\textbf{H}}]_{uu}+\Delta\to[\boldsymbol{\Sigma}^{-1}_{\textbf{H}}]_{uu},\;\text{as}\;N_{\mathsf{t}}\to\infty.
\end{align}
\end{IEEEproof}

\subsection{Ergodic Achievable Rate of UAV-Eve}
The ergodic achievable rate of the UAV-Eve is written from \eqref{eq:SINR_eve}, \eqref{eq:secrecy_rate} as follows: 
\begin{align}\label{eq:rate_eve}
    \mathsf{R}_{\mathsf{e}} &=\mathbb{E}\left\{\log_{2}\left(1+\frac{\phi(1-\delta)|\textbf{h}_\mathsf{e}^{\rm H}\textbf{w}_u|^2}{\sum_{i=1}^{N_{\mathsf{AN}}} (1-\phi)|\textbf{h}_\mathsf{e}^{\rm H}\textbf{v}_i|^2 + \rho^{{-}1}_\mathsf{e}}\right)\right\}\nonumber\\
    &\approx\log_{2}\left(1+\frac{\phi(1-\delta)\mathbb{E}\left\{|\textbf{h}_\mathsf{e}^{\rm H}\textbf{w}_u|^2\right\}}{\sum_{i=1}^{N_{\mathsf{AN}}} (1-\phi)\mathbb{E}\left\{|\textbf{h}_\mathsf{e}^{\rm H}\textbf{v}_i|^2\right\} + \rho^{{-}1}_\mathsf{e}}\right),
\end{align}
where the approximation comes from \cite[Lemma 1]{zhang2014power}. We derive the closed-form expressions for both the conventional and the proposed precoder designs from above equation.
\subsubsection{Conventional precoder design}
\begin{theorem}
The closed-form expression of the ergodic achievable rate of UAV-Eve for the conventional precoder design is given by
\begin{align}\label{eq:theo3:def}
&\mathsf{R}_{\mathsf{e}}^{\mathsf{cv}}\nonumber\\ &=\log_{2}\left(1+\frac{\phi(1-\delta)\left(\frac{1}{K(\kappa+1)}+\frac{\kappa}{\kappa+1}\text{Tr}\left[\textbf{h}_{\mathsf{LoS},\rm e}\textbf{h}_{\mathsf{LoS},\rm e}^{\rm H}\Gamma_{\textbf{w}}\right]\right)}{(1-\phi)\left(\frac{\kappa}{\kappa+1}\text{Tr}\left[\textbf{h}_{\mathsf{LoS},\rm e}\textbf{h}_{\mathsf{LoS},\rm e}^{\rm H}\Gamma_{\textbf{V}}\right]+\frac{1}{\kappa+1}\right) + \rho^{{-}1}_\mathsf{e}}\right).
\end{align}
\end{theorem}
\begin{IEEEproof}
See Appendix C.
\end{IEEEproof}
\subsubsection{Proposed precoder design}
In this part, we derive the closed-form expression of the ergodic achievable rate of UAV-Eve with the assumption of that $\hat{\theta}_{\rm e,\mathsf{LoS}}=\theta_{\rm e,\mathsf{LoS}}$, which means that the GS knows the exact angle direction of the UAV-Eve. Instead, we discuss the effect of the angular calibration error in the later subsection separately.
\begin{theorem}
The closed-form expression of the ergodic achievable rate of UAV-Eve for the proposed precoder design is expressed as
\begin{align}\label{eq:theo4:def}
\mathsf{R}_{\mathsf{e}}^{\mathsf{pp}} &=\log_{2}\left(1+\frac{\phi(1-\delta)\left(\frac{1}{K(\kappa+1)}\right)}{(1-\phi)\left(\frac{ \kappa }{ \kappa + 1 }\frac{N_{\mathsf{t}}-K-1}{[\boldsymbol{\Sigma}^{-1}_{\textbf{G}}]_{\mathsf{ee}}}+\frac{ 1 }{ \kappa + 1 }\right)+\rho^{{-}1}_\mathsf{e}}\right).
\end{align}
\end{theorem}
\begin{IEEEproof}
See Appendix D.
\end{IEEEproof}

Since the ZF precoding in the proposed precoder cancels out  data signal from the LoS channel but focuses AN power to the LoS channel, we can make sure that the ergodic achievable rate of the UAV-Eve for the proposed precoder design is lower than the conventional precoder design as
\begin{align}\label{eq:comp_rate:eve}
    \mathsf{R}_{\mathsf{e}}^{\mathsf{cv}} > \mathsf{R}_{\mathsf{e}}^{\mathsf{pp}}.
\end{align}
However, it is not tractable to show it directly by the derived expressions in \textit{Theorem 3} and \textit{Theorem 4}. Instead, we show that \eqref{eq:comp_rate:eve} asymptotically holds as $N_{\mathsf{t}}$ goes infinity.
\begin{corollary}
If $N_{\mathsf{t}}\to\infty$, $   \mathsf{R}_{\mathsf{e}}^{\mathsf{cv}} > \mathsf{R}_{\mathsf{e}}^{\mathsf{pp}}.$
\end{corollary}
\begin{IEEEproof}
We first obtain
\begin{align}
    \Gamma_{\textbf{V}}&=\mathbb{E}\left\{\textbf{V}^{\mathsf{cv}}\left(\textbf{V}^{\mathsf{cv}}\right)^{\rm H}\right\}\nonumber\\
    &\overset{(a)}{=}\mathbb{E}\left\{\frac{1}{N_{\mathsf{AN}}}\left(\textbf{I}-\textbf{H}\left(\textbf{H}^{\rm H}\textbf{H}\right)^{-1}\textbf{H}^{\rm H}\right)\right\}\nonumber\\
    &=\frac{1}{N_{\mathsf{t}}-K}\left(\textbf{I}-\mathbb{E}\left\{\textbf{H}\left(\textbf{H}^{\rm H}\textbf{H}\right)^{-1}\textbf{H}^{\rm H}\right)\right\}
\end{align}
where (a) comes from \cite{zhu2015linear}. Then, by applying the property, $\frac{1}{N_{\mathsf{t}}}\textbf{H}^{\rm H}\textbf{H}\to\textbf{I}$ as $N_{\mathsf{t}}$ goes infinity \cite[Lemma 2]{zhang2014power}, we can rewrite
\begin{align}
    &\text{Tr}\left[\textbf{h}_{\mathsf{LoS},\rm e}\textbf{h}_{\mathsf{LoS},\rm e}^{\rm H}\Gamma_{\textbf{V}}\right]\nonumber\\
    &\to\frac{1}{N_{\mathsf{t}}-K}\text{Tr}\left[\textbf{h}_{\mathsf{LoS},\rm e}\textbf{h}_{\mathsf{LoS},\rm e}^{\rm H}\left(\textbf{I}-\mathbb{E}\left\{\frac{1}{N_{\mathsf{t}}}\textbf{H}\textbf{H}^{\rm H}\right\}\right)\right]\nonumber\\
    &=\frac{1}{N_{\mathsf{t}}-K}\text{Tr}\left[\textbf{h}_{\mathsf{LoS},\rm e}\textbf{h}_{\mathsf{LoS},\rm e}^{\rm H}\left(\textbf{I}-\frac{1}{N_{\mathsf{t}}}\left(\frac{K}{\kappa+1}\textbf{I}\right.\right.\right.\nonumber\\
    &\left.\left.\left.+\frac{\kappa}{\kappa+1}\textbf{H}_{\mathsf{LoS}}\textbf{H}_{\mathsf{LoS}}^{\rm H}\right)\right)\right]\nonumber\\
    &\to\frac{1}{N_{\mathsf{t}}-K}\text{Tr}\left[\textbf{h}_{\mathsf{LoS},\rm e}\textbf{h}_{\mathsf{LoS},\rm e}^{\rm H}\right]=\frac{N_{\mathsf{t}}}{N_{\mathsf{t}}-K}\to1.
\end{align}
By plugging it in \eqref{eq:theo3:def}, if $N_{\mathsf{t}}\to\infty$, we can get 
\begin{align}\label{eq:col3:p3}
    &\mathsf{R}_{\mathsf{e}}^{\mathsf{cv}}\nonumber\\ &\to\log_{2}\left(1+\frac{\phi(1-\delta)\left(\frac{1}{K(\kappa+1)}+\frac{\kappa}{\kappa+1}\text{Tr}\left[\textbf{h}_{\mathsf{LoS},\rm e}\textbf{h}_{\mathsf{LoS},\rm e}^{\rm H}\Gamma_{\textbf{w}}\right]\right)}{(1-\phi) + \rho^{{-}1}_\mathsf{e}}\right).
\end{align}
We can also easily get
\begin{align}
\frac{N_{\mathsf{t}}-K-1}{[\boldsymbol{\Sigma}^{-1}_{\textbf{G}}]_{\mathsf{ee}}}>1,\;\text{as}\;N_{\mathsf{t}}\to\infty.
\end{align}
By applying it to \eqref{eq:theo4:def} and compare it with \eqref{eq:col3:p3}, we can obtain the result.
\end{IEEEproof}

Now, we can obtain one of the key results by combining results from \textit{Corollary 2} and \textit{Corollary 3}.
\begin{corollary}
The ergodic secrecy rate for the proposed precoder design is always greater than the conventional precoder design as $N_{\mathsf{t}}$ grows without bound as follow
\begin{align}\label{}
    \mathsf{R}^\mathsf{sec,cv}_u < \mathsf{R}^\mathsf{sec,pp}_u,\;\text{as}\;N_{\mathsf{t}}\to\infty,
\end{align}
where $\mathsf{R}^\mathsf{sec,cv}_u$, $\mathsf{R}^\mathsf{sec,pp}_u$ is the ergodic secrecy rate for the conventional precoder and the proposed precoder design, respectively.
\end{corollary}

We can also observe another interesting asymptotic behavior of the ergodic achievable rate of the UAV-Eve when Rician K-factor $\kappa$ grows without bound. As $\kappa\to\infty$,  the result in \textit{Theorem 3} tends to
\begin{align}\label{}
&\mathsf{R}_{\mathsf{e}}^{\mathsf{cv}}\nonumber\\ &\to\log_{2}\left(1+\frac{\phi(1-\delta)\left(\text{Tr}\left[\textbf{h}_{\mathsf{LoS},\rm e}\textbf{h}_{\mathsf{LoS},\rm e}^{\rm H}\Gamma_{\textbf{w}}\right]\right)}{(1-\phi)\left(\text{Tr}\left[\textbf{h}_{\mathsf{LoS},\rm e}\textbf{h}_{\mathsf{LoS},\rm e}^{\rm H}\Gamma_{\textbf{V}}\right]\right) + \rho^{{-}1}_\mathsf{e}}\right).
\end{align}
On the other hand, the result in \textit{Theorem 4} converges to
\begin{align}\label{}
&\mathsf{R}_{\mathsf{e}}^{\mathsf{pp}}\nonumber\\
&=\log_{2}\left(1+\frac{\phi(1-\delta)}{K(1-\phi)\left(\frac{\kappa (N_{\mathsf{t}}-K-1)}{[\boldsymbol{\hat{\Sigma}}^{-1}_{\textbf{G}}]_{\mathsf{ee}}}+1\right)+K(\kappa+1)\rho^{{-}1}_\mathsf{e}}\right)\nonumber\\
&\to0.
\end{align}
Note that this behavior holds regardless of $\phi$ (e.g. $\phi=1$). From the above observation, we can also have the following corollary.
\begin{corollary}
The ergodic achievable rate of the UAV-Eve for the proposed precoder design goes to zero as $\kappa\to\infty$, which means that we can completely protect the data from the UAV-Eve and achieve the perfect security.
\end{corollary}
\subsection{Power Splitting Factor ($\phi$) Allocation Strategy for the Proposed Precoder Design}
In this subsection, we optimize the power splitting factor ($\phi$) for the proposed precoder design by the observation of the asymptotic behavior of the ergodic achievable rate. The power splitting factor decides the balance between the transmitted power of data and AN. If we allocate more power to the AN, we can enhance the protection of the data from the UAV-Eve but the data rate of UAV-UEs decreases. Thus, it is meaningful to find the optimal value. We first replace $\phi$ with $1-\frac{1}{N_{\mathsf{t}}^{\alpha}}$. Note that $0\leq\phi\leq1$ holds if $\alpha$ is non-negative real-value. Then, we present the following theorem from the result of \textit{Theorem~4}.
\begin{theorem}
If we allocate the $0\leq\alpha<1$ in $\phi=1-\frac{1}{N_{\mathsf{t}}^{\alpha}}$, we have
\begin{align}\label{eq:theo5:def}
    \mathsf{R}_{\mathsf{e}}^{\mathsf{pp}}&\to0,\;\text{as}\;N_{\mathsf{t}}\to\infty.
\end{align}
\end{theorem}
\begin{IEEEproof}
Substituting $\phi=1-\frac{1}{N_{\mathsf{t}}^{\alpha}}$ into \eqref{eq:theo4:def}, we get
\begin{align}\label{}
\mathsf{R}_{\mathsf{e}}^{\mathsf{pp}} &=\log_{2}\left(1+\frac{\left(1-\frac{1}{N_{\mathsf{t}}^{\alpha}}\right)(1-\delta)\left(\frac{1}{K(\kappa+1)}\right)}{\frac{1}{N_{\mathsf{t}}^{\alpha}}\left(\frac{ \kappa }{ \kappa + 1 }\frac{N_{\mathsf{t}}-K-1}{[\boldsymbol{\Sigma}^{-1}_{\textbf{G}}]_{\mathsf{ee}}}+\frac{ 1 }{ \kappa + 1 }\right)+\rho^{{-}1}_\mathsf{e}}\right).
\end{align}
When, $0\leq\alpha<1$, we can easily obtain \eqref{eq:theo5:def}.
\end{IEEEproof}
Note that when $\alpha=1$, \textit{Theorem 4} converges to
\begin{align}\label{}
\mathsf{R}_{\mathsf{e}}^{\mathsf{pp}} &\to\log_{2}\left(1+\frac{(1-\delta)\left(\frac{1}{K(\kappa+1)}\right)}{\left(\frac{ \kappa }{ \kappa + 1 }\right)+\rho^{{-}1}_\mathsf{e}}\right),\;\text{as}\;N_{\mathsf{t}}\to\infty.
\end{align}
When $\alpha>1$, we can get
\begin{align}\label{eq:theo5:p3}
\mathsf{R}_{\mathsf{e}}^{\mathsf{pp}} &\to\log_{2}\left(1+\frac{(1-\delta)\left(\frac{1}{K(\kappa+1)}\right)}{\rho^{{-}1}_\mathsf{e}}\right),\;\text{as}\;N_{\mathsf{t}}\to\infty.
\end{align}

\textit{Theorem 5} shows that if we allocate $0\leq\alpha<1$. it means that we already put too much power on AN since the ergodic achievable rate of UAV-Eve is zero. On the other hand, if we allocate $\alpha>1$, it means that we need to reduce AN power since the ergodic achievable rate of UAV-Eve already reaches the maximum point. As a result, we have the following Corollary.
\begin{corollary}
For the proposed precoder design, when $N_{\mathsf{t}}$ grows without bound and if the maximum value exists $0\leq\phi<1$, the optimal value that maximizes the secrecy rate is $\alpha=1$, which indicates $\phi=1-\frac{1}{N_{\mathsf{t}}}$.
\end{corollary}

Note that if the maximum ergodic achievable rate of UAV-Eve in \eqref{eq:theo5:p3} is not sufficiently high so that the secrecy rate becomes the increasing function, then we do not have to allocate any power to the AN and the secrecy rate is maximized at $\phi=1$.

\subsection{Effect of Angular Calibration Error $\epsilon$}
In this subsection, we discuss the effect of angular calibration error ($\epsilon$) on the proposed precoder design. In previous sections, we assume that the GS utilizes the perfect elevation angle of the UAV-Eve for the proposed precoder design and derives the ergodic achievable rates. However, imperfect angle information leads to leakage in the ZF precoding.

From \eqref{eq:LoS_angle}, we can express the LoS component of the small-scale fading channel of the UAV-Eve as
\begin{align}
    \textbf{h}_{\mathsf{LoS},\mathsf{e}}&=\textbf{a}_{N_{\mathsf{t}}}(\hat{\theta}_{\mathsf{e},\mathsf{LoS}}-\epsilon).
\end{align}
By \eqref{eq:steering_vector}, we can rewrite the entry of the vector as
\begin{align}
    \left[\textbf{h}_{\mathsf{LoS},\mathsf{e}}\right]_{n}&=e^{-j\frac{2\pi d_{\rm s}}{\lambda}(n-1)\sin\left(\hat{\theta}_{\mathsf{e},\mathsf{LoS}}-\epsilon\right)}
\end{align}
where $[\textbf{x}]_{n}$ indicates $n_{\rm th}$ entry of the vector. By using $1_{\mathsf{st}}$ order Taylor expansion at $\epsilon=0$, we can approximate it as
\begin{align}
    &\left[\textbf{h}_{\mathsf{LoS},\mathsf{e}}\right]_{n}\approx e^{-j\frac{2\pi d_{\rm s}}{\lambda}(n-1)\sin\left(\hat{\theta}_{\mathsf{e},\mathsf{LoS}}\right)}\nonumber\\
    &+\epsilon\cos(\hat{\theta}_{\mathsf{e},\mathsf{LoS}})\frac{2\pi d_{\rm s}}{\lambda}(n-1)e^{-j\left(\frac{2\pi d_{\rm s}}{\lambda}(n-1)\sin\left(\hat{\theta}_{\mathsf{e},\mathsf{LoS}}\right)-\frac{\pi}{2}\right)}\nonumber\\
    &=\left[\hat{\textbf{h}}_{\mathsf{LoS},\mathsf{e}}\right]_{n}\nonumber\\
    &+\epsilon\cos(\hat{\theta}_{\mathsf{e},\mathsf{LoS}})\frac{2\pi d_{\rm s}}{\lambda}(n-1)e^{-j\left(\frac{2\pi d_{\rm s}}{\lambda}(n-1)\sin\left(\hat{\theta}_{\mathsf{e},\mathsf{LoS}}\right)-\frac{\pi}{2}\right)}~.
\end{align}
Then, we can obtain the MSE of the estimated LoS component channel as
\begin{align}\label{eq:MSE_ang_cal}
    &\mathbb{E}\left\{\|\textbf{h}_{\mathsf{LoS},\mathsf{e}}-\hat{\textbf{h}}_{\mathsf{LoS},\mathsf{e}}\|^2\right\}\nonumber\\
    &=\sigma_{\epsilon}^2\cos^2(\hat{\theta}_{\mathsf{e},\mathsf{LoS}})\sum_{n=1}^{N_{\mathsf{t}}}\frac{4\pi^2 (d_{\rm s})^2}{\lambda^2}(n-1)^2\nonumber\\
    &=\sigma_{\epsilon}^2\cos^2(\hat{\theta}_{\mathsf{e},\mathsf{LoS}})\frac{4\pi^2 (d_{\rm s})^2N_{\mathsf{t}}(N_{\mathsf{t}}-1)(2N_{\mathsf{t}}-1)}{6\lambda^2}
\end{align}
From \eqref{eq:MSE_ang_cal}, we can conclude that the performance degradation by the angular calibration error depends on the variance of error ($\sigma_{\epsilon}^2$), the number of GS antennas ($N_{\mathsf{t}}$), and the elevation angle ($\theta_{\mathsf{e},\mathsf{LoS}}$). The degradation increases when $\sigma_{\epsilon}^2$, $N_{\mathsf{t}}$ increase and $\theta_{\mathsf{e},\mathsf{LoS}}$ decrease, which correspond to UAV-Eve being far from the GS if the height of UAV-Eve is fixed.

\section{Fingerprint Embedding Authentication}

In this section, we introduce the fingerprint embedding authentication framework. Authentication is a different aspect of the physical layer security than the secrecy rate that we discuss in the previous section. Thus, it has different performance metrics, which calculate the probability of whether the eavesdropper is able to recover the secret key for the authentication and impersonate the legitimate user. The fingerprint embedding authentication framework has been studied in point-to-point communications \cite{paul2011mimo,perazzone2021artificial}, but it is not extended to the multi-user MIMO scenario and the performance comparison between different precoder designs is not focused on. In this paper, we apply the fingerprint embedding framework in our scenario described in Section~\ref{sec:system}, along with the conventional and the proposed precoder designs.

The authentication tag is generated from the GS by a one-way collision-resistant function (e.g. cryptographic hash function) as follows: 
\begin{align}
    \textbf{t}_k&=f(\textbf{s}_k,\eta_k),
\end{align}
where $\eta_k$ is a secret key that is shared between the GS and the $k_{\mathsf{th}}$ UAV-UE. The GS superimposes the authentication tags on the data symbols ($s_k$) with a fraction of the total power using the tag power factor ($\delta$) such that the data rate is not affected, and it transmits them to the UAV-UEs. It is worth noting that data symbols and tags are precoded by $\textbf{W}$ before being transmitted.

\subsection{Authentication Procedure of the UAV-UE}

The UAV-UE estimates authentication tag from the received signal in \eqref{eq:received_signal:user} as follows
\begin{align}
    \textbf{r}_u^{\rm H}&=\textbf{y}_u^{\rm H}-\frac{\sqrt{\mathsf{P}_\mathsf{Tx}}}{\sqrt{\mathsf{PL}_u}}\textbf{h}_u^{\rm H}\sqrt{\phi}\textbf{w}_k(\sqrt{(1-\delta)}\textbf{s}_k)^{\rm H}
\end{align}
where $\textbf{r}_u$ denotes residual signal. Then, the estimated authentication tag can be expressed as 
\begin{align}
    \hat{\textbf{t}}_u^{\rm H}&=\frac{\sqrt{\mathsf{PL}_{u}}\textbf{r}_u^{\rm H}}{\sqrt{\mathsf{P}_\mathsf{Tx}\delta\phi}\textbf{h}_u^{\rm H}\textbf{w}_u}\nonumber\\
    &=\textbf{t}_u^{\rm H}+\underbrace{\sum_{k\neq u}\frac{\textbf{h}_u^{\rm H}\textbf{w}_k}{\sqrt{\delta}\textbf{h}_u^{\rm H}\textbf{w}_u}(\sqrt{1-\delta}\textbf{s}_k+\sqrt{\delta}\textbf{t}_k)^{\rm H}}_\text{multiuser interference}\nonumber\\
    &+\underbrace{\sum_{i=1}^{N_{\mathsf{AN}}}\frac{\sqrt{1-\phi}\textbf{h}_u^{\rm H}\textbf{v}_i}{\sqrt{\delta\phi}\textbf{h}_u^{\rm H}\textbf{w}_u}\textbf{z}_i^{\rm H}}_\text{artificial noise}+\underbrace{\frac{\sqrt{\mathsf{PL}_{u}}}{\sqrt{\mathsf{P}_\mathsf{Tx}\delta\phi}\textbf{h}_u^{\rm H}\textbf{w}_u}\textbf{n}_u^{\rm H}}_\text{additive noise}.
\end{align}
We assume that the UAV-UE successfully decodes data symbols without errors and substrates the corresponding contributions from the received signal, and then equalizes it. Since both the conventional precoders and the proposed precoders are designed such that they eliminate the multi-user interference and the artificial noise, we can rewrite the estimated authentication tag as
\begin{align}
    \hat{\textbf{t}}_u^{\rm H}&=\textbf{t}_u^{\rm H}+\frac{\sqrt{\mathsf{PL}_{u}}}{\sqrt{\mathsf{P}_\mathsf{Tx}\delta\phi}\textbf{h}_u^{\rm H}\textbf{w}_u}\textbf{n}_u^{\rm H}.
\end{align}
Next, the UAV-UE generates the expected tag by using already known shared key and decoded data symbols, as
\begin{align}
    \Tilde{\textbf{t}}_u&=f(\hat{\textbf{s}}_u,\eta_u),
\end{align}
where $f(\cdot)$ indicates tag-generation function which is commonly implemented by a cryptographic hash function.
Then, a binary hypothesis test is carried out to authenticate the message. The test statistic is given by the correlation between the expected tag and the estimated tag, as
\begin{align}
    \tau_b&=\Re(\hat{\textbf{t}}_u^{\rm H}\Tilde{\textbf{t}}_u),
\end{align}
where $\Re(\cdot)$ refers to a function return the real part of the complex argument. The threshold hypotheses test is designed by
\begin{align*}
    H_0:& \text{ not authentic,}\quad\tau_b\leq\tau_{\mathsf{thr}},\\
    H_1:& \text{ authentic,} \quad\tau_b>\tau_{\mathsf{thr}},
\end{align*}
where $\tau_{\mathsf{thr}}$ is the test threshold. If the channel matrix is regarded as a deterministic matrix, the mean and the variance of $\tau_b$ under each hypotheses is given by
\begin{align}
    \mathbb{E}\{\tau_b|H_0\}&=\mu_{u,0}=0,\label{eq:hypo_mean0:UE}\\
    \text{Var}\{\tau_b|H_0\}&=\nu_{u,0}^2=\frac{L_{\rm t}}{2}\left(1+\frac{\rho^{-1}_u}{\delta\phi|\textbf{h}_u^{\rm H}\textbf{w}_u|^2}\right),\label{eq:hypo_var0:UE}\\
    \mathbb{E}\{\tau_b|H_1\}&=\mu_{u,1}=L_{\rm t},\label{eq:hypo_mean1:UE}\\
    \text{Var}\{\tau_b|H_1\}&=\nu_{u,1}^2=\text{Var}(\|\textbf{t}_u\|^2)+\frac{L_{\rm t}}{2}\left(\frac{\rho^{-1}_u}{\delta\phi|\textbf{h}_u^{\rm H}\textbf{w}_u|^2}\right)\label{eq:hypo_var1:UE},
\end{align}
where $L_{\mathsf{t}}$ is the length of the authentication tag and we assume that the tag and data are uncorrelated. If we consider complex-valued Gaussian distribution for the tag, $\text{Var}(\|\textbf{t}_u\|^2)=L_{\mathsf{t}}$. With sufficiently large $L_t$, we can approximate the distribution of $\tau_b$ as Gaussian by the central limit theorem, and we can obtain the threshold that limits the probability that the authentication accepts the wrong tag as follows
\begin{align}\label{eq:tau_thr}
    \tau_{\mathsf{thr}}&=\Phi^{-1}(1-p_{\rm fa})\nu_{u,0},
\end{align}
where $\Phi(\cdot)$ is the CDF of the Gaussian distribution, and $p_{\rm fa}$ denotes a false alarm probability that admits the wrong authentication tag. To the end, the authentication  probability that the true tag is accepted can be expressed as
\begin{align}\label{eq:P_A}
        P_{\rm A}&=1-\Phi\left(\frac{\tau_{\mathsf{thr}}-\mu_{u,1}}{\nu_{u,1}}\right).
\end{align}
\subsection{Impersonation Procedure of the UAV-Eve}
The UAV-Eve estimates the authentication tag from the received signal in \eqref{eq:received_signal:Eve} as
\begin{align}
    \hat{\textbf{t}}_\mathsf{e}^{\rm H}&=\textbf{t}_u^{\rm H}+\underbrace{\sum_{i=1}^{N_{\mathsf{AN}}}\frac{\sqrt{1-\phi}\textbf{h}_\mathsf{e}^{\rm H}\textbf{v}_i}{\sqrt{\delta\phi}\textbf{h}_\mathsf{e}^{\rm H}\textbf{w}_u}\textbf{z}_i^{\rm H}}_\text{artificial noise}+\underbrace{\frac{\sqrt{\mathsf{PL}_{\mathsf{e}}}}{\sqrt{\mathsf{P}_\mathsf{Tx}\delta\phi}\textbf{h}_\mathsf{e}^{\rm H}\textbf{w}_u}\textbf{n}_{\mathsf{e}}^{\rm H}}_\text{additive noise}.
\end{align}
We would like to remark that the worst-case assumption is considered in the above equation, in which the UAV-Eve is able to fully eliminate the contributions from messages and the tags of the other UAV-UEs. Next, the UAV-Eve generates the expected authentication tag by the decoded messages. Since the UAV-Eve does not know the secret key, it generates all possible tags by iterative keys selection from the keyspace $|\mathcal{K}|$, and find the best key that maximizes the correlation with the estimated tag. In other words, UAV-Eve faces $|\mathcal{K}|$ hypotheses tests. The test statistic is given by
\begin{align}
    \tau_{\rm e}&=\Re(\hat{\textbf{t}}_{\rm e}^{\rm H}\Tilde{\textbf{t}}_i).
\end{align}
The best key is decided by the maximum likelihood (ML) estimation as
\begin{align}
   \eta_i&=\arg\max_{\eta\in\mathcal{K}}\tau_{\rm e}(\eta).
\end{align}
Note that we do not take into account the computation complexity of the ML estimator in searching all possible keys. The mean and the variance of $\tau_{\rm e}$ under the $H_0$, $H_1$ hypotheses are expressed as
\begin{align}
    &\mathbb{E}\{\tau_{\rm e}|H_0\}=\mu_{\rm e,0}=0,\\
    &\text{Var}\{\tau_{\mathsf{e}}|H_0\}=\nu_{\mathsf{e},0}^2\nonumber\\
    &=\frac{L_{\rm t}}{2}\left(1+\sum_{i=1}^{N_{\mathsf{AN}}}\frac{(1-\phi)|\textbf{h}_\mathsf{e}^{\rm H}\textbf{v}_i|^2}{\delta\phi|\textbf{h}_\mathsf{e}^{\rm H}\textbf{w}_u|^2}+\frac{\rho^{-1}_\mathsf{e}}{\delta\phi|\textbf{h}_\mathsf{e}^{\rm H}\textbf{w}_u|^2}\right)\label{eq:hypo_var0:Eve},
\end{align}
\begin{align}
    &\mathbb{E}\{\tau_{\rm e}|H_1\}=\mu_{\rm e, 1}=L_{\rm t},\\
    &\text{Var}\{\tau_\mathsf{e}|H_1\}=\nu_{\mathsf{e},1}^2\nonumber\\
    &=\text{Var}(\|\textbf{t}_u\|^2)+\frac{L_{\rm t}}{2}\left(\sum_{i=1}^{N_{\mathsf{AN}}}\frac{(1-\phi)|\textbf{h}_\mathsf{e}^{\rm H}\textbf{v}_i|^2}{\delta\phi|\textbf{h}_\mathsf{e}^{\rm H}\textbf{w}_u|^2}+\frac{\rho^{-1}_\mathsf{e}}{\delta\phi|\textbf{h}_\mathsf{e}^{\rm H}\textbf{w}_u|^2}\right)\label{eq:hypo_var1:Eve}.
\end{align}
Then, we can derive the probability that the UAV-Eve successfully recovers the secret key in a single observation by the ML estimator \cite{perazzone2021artificial}, as
\begin{align}
    P_{K}&=\int^{\infty}_{-\infty}\Phi\left(\frac{\tau-\mu_{\rm e,0}}{\nu_{\rm e,0}}\right)^{|\mathcal{K}|-1}\varphi\left(\frac{\tau-\mu_{\rm e,1}}{\nu_{\rm e,1}}\right){\rm d}\tau,\label{eq:P_K}
\end{align}
where $\varphi(\cdot)$ is the probability density function (PDF) of Gaussian distribution.

\subsection{Tag Power Factor ($\delta$) Allocation Strategy}
In this subsection, we design an algorithm to optimize $\delta$. The GS needs to prevent the UAV-Eve from recovering the secret key, while the UAV-UE surely authenticates the messages with high probability. In this sense, the problem of interest is given by
\begin{IEEEeqnarray}{rl}
    \min_{\delta}
    &\quad P_{\mathsf{K}}, \label{eq:optimization_ssr_1}\\
    \text{s.t.}
    &\quad P_{\mathsf{A}} \geq p_{\mathsf{thr}}, \IEEEyessubnumber \label{eq:optimization_ssr_2}
\end{IEEEeqnarray}
where $p_{\mathsf{thr}}$ is the threshold of the authentication probability. From the variances of the hypothesis test statistics in \eqref{eq:hypo_var0:UE}, \eqref{eq:hypo_var1:UE}, \eqref{eq:hypo_var0:Eve}, \eqref{eq:hypo_var1:Eve}, we can easily observe that all variances are decreasing functions of $\delta$, which means that both $P_{\mathsf{A}}$ and  $P_{\mathsf{K}}$ are increasing function of $\delta$. Therefore, it is concluded that $\delta$ reaches the optimized point when $P_{\mathsf{A}}(\delta_{\mathsf{opt}}) = p_{\mathsf{thr}}$. By using the fact that $P_{\mathsf{A}}(\delta)$ is an increasing function starting from 0 to 1, we can find $\delta_{\mathsf{opt}}$ by an iterative algorithm as illustrated in Algorithm~1. 

Furthermore, we also consider the dependency of the power splitting factor ($\phi$) on the authentication probability in the algorithm. It is worth noting that the power splitting factor divides power between data plus tag and artificial noise, and high $\phi$ leads to high power allocation of the authentication tag.
We define a new constant, as
\begin{align}
    \psi &= \delta\phi,
\end{align}
where $\psi$ decides the joint contribution of two power factors in the authentciation probability.
Then, we can rewrite the variances of the test statistic in \eqref{eq:hypo_var0:UE}, \eqref{eq:hypo_var1:UE} as
\begin{align}
        \nu_{u,0}^2&=\frac{L_{\rm t}}{2}\left(1+\frac{\rho^{-1}_u}{\psi|\textbf{h}_u^{\rm H}\textbf{w}_u|^2}\right),\label{eq:hypo_var00:UE}\\
        \nu_{u,1}^2&=\text{Var}(\|\textbf{t}_u\|^2)+\frac{L_{\rm t}}{2}\left(\frac{\rho^{-1}_u}{\psi|\textbf{h}_u^{\rm H}\textbf{w}_u|^2}\right)\label{eq:hypo_var11:UE}.
\end{align}
Then, we can find the optimal $\psi$ that satisfies $P_{\mathsf{A}}(\psi) = p_{\mathsf{thr}}$. By finding the optimal $\psi$ first, we can easily find the $\delta_{\mathsf{opt}}$ even if we vary $\phi$.
\begin{algorithm}
  \caption{Iterative algorithm for finding the optimal $\delta$}
  \label{algorithm:overall}
    \begin{algorithmic}[1]
        \State \textbf{Initialize:} $k \gets 1$, $P_{\mathsf{A}}^{(k-1)} \gets 0$, $\psi^{(k-1)} \gets 0$
        \While{$ P_{\mathsf{A}}^{(k-1)} < p_{\mathsf{thr}}$} \label{algorithm:convergence_start}
            \State Compute $\psi^{(k)} \gets \psi^{(k-1)} + \psi_{\Delta}$
            \State Compute  $P_{\mathsf{A}}^{(k)}$ by \eqref{eq:hypo_mean1:UE}, \eqref{eq:tau_thr}, \eqref{eq:P_A}, \eqref{eq:hypo_var00:UE}, \eqref{eq:hypo_var11:UE}
            \State $k \gets k+1$
        \EndWhile
        \State $\delta_{\mathsf{opt}}=\frac{\psi^{(k-1)}}{\phi}$
        \label{algorithm:convergence_end}
    \end{algorithmic}
\end{algorithm}

\section{Numerical Results} \label{sec:results}
In this section, we present simulation results for the performance of the proposed precoder design by comparing it with the conventional precoder design. We consider a scenario that is an extension of Fig.~\ref{fig:illu}, and the specific simulation settings/parameters are listed in Table~\ref{table:settings}.

\begin{table}[!t]
\renewcommand{\arraystretch}{1.1}
\caption{Simulation settings}
\label{table:settings}
\centering
\begin{tabular}{lc}
\hline
Parameter & Value \\
\hline\hline
Transmit power ($\mathsf{P}_\mathsf{Tx}$) & $[25, 30, 35]$~dBm\\
Number of antennas at UAV-BS ($N_{\mathsf{t}}$) & $[16, 128]$ \\
Number of users ($K$) & $4$ \\
Number of UAV-Eve & $1$ \\
K-factor of Rician channel ($\kappa$) & $[10, 30]$~dB \\
UAV-UEs horizontal distance to GS distribution ($d_{u}$) & $\mathcal{U}[10,100]$ m \\
UAV-Eve horizontal distance to GS ($d_{\mathsf{e}}$) & $10$ m \\
UAVs height ($h_{\mathsf{UAV}}$) & $100$ m \\
Angular calibration error ($\sigma_{\epsilon}$) & $[0, 5, 10]$ $^{\circ}$\\
Length of authentication tag ($L_{\rm t}$) & $1024$ \\
Number of key bits & $64$ \\
False alarm probability ($p_{\rm fa}$) & $0.001$\\
Threshold authentication probability ($p_{\rm thr}$) & $0.999$\\
Thermal noise & $-174$~dBm/Hz\\
Noise figure & $9$~dB\\\hline
\hline
\end{tabular}
\vspace{-0.15in}
\end{table}

\subsection{Ergodic Sum Secrecy Rate versus Transmit Power}
\begin{figure}[t]
	\centering
	\subfloat[$N_{\mathsf{t}}=16$]{
\includegraphics[width=0.5\textwidth]{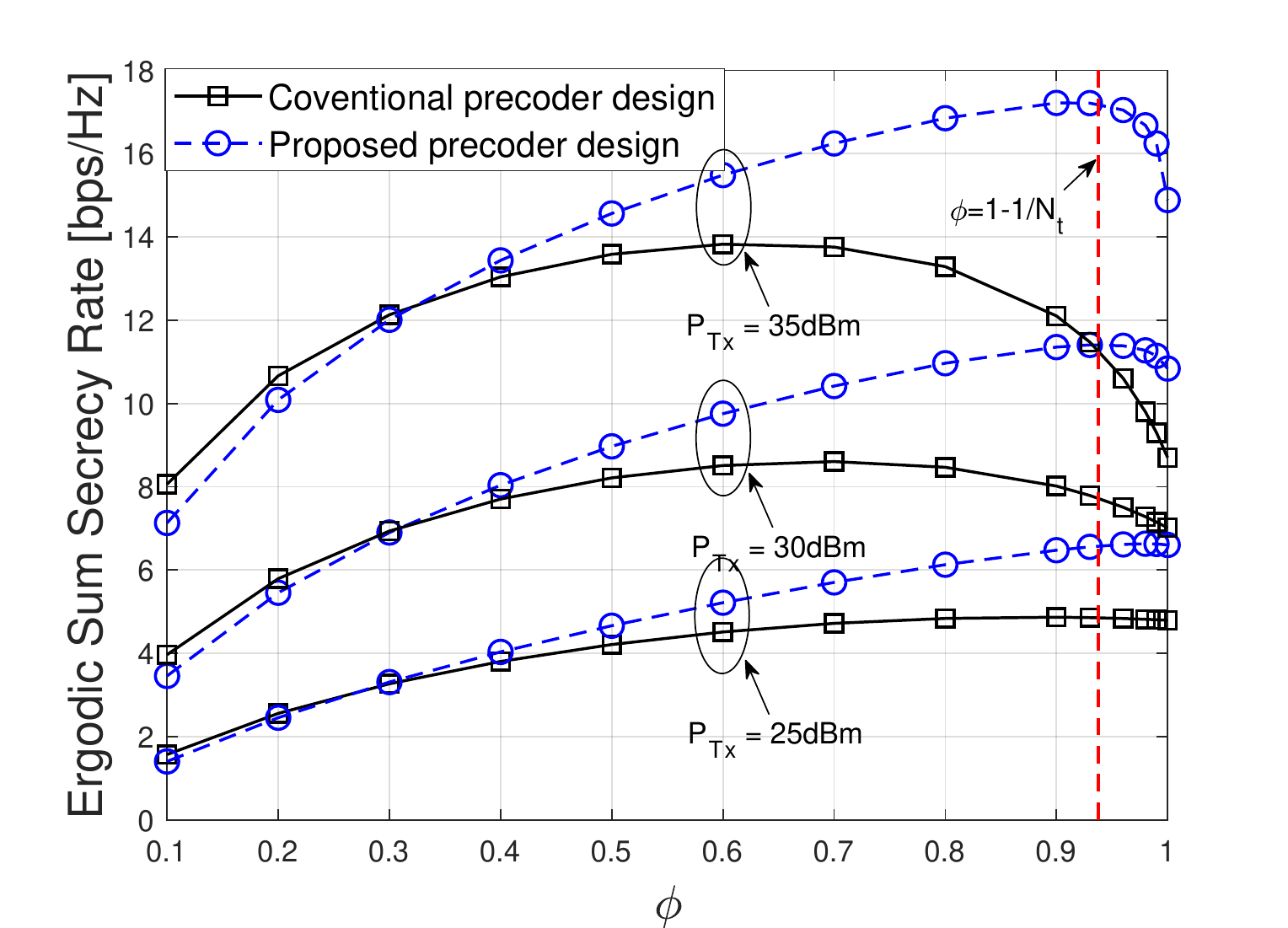}\label{fig:SR_Ptx_Nt16}}
\vspace{-0.01in}
	\subfloat[$N_{\mathsf{t}}=128$]{
	\includegraphics[width=0.5\textwidth]{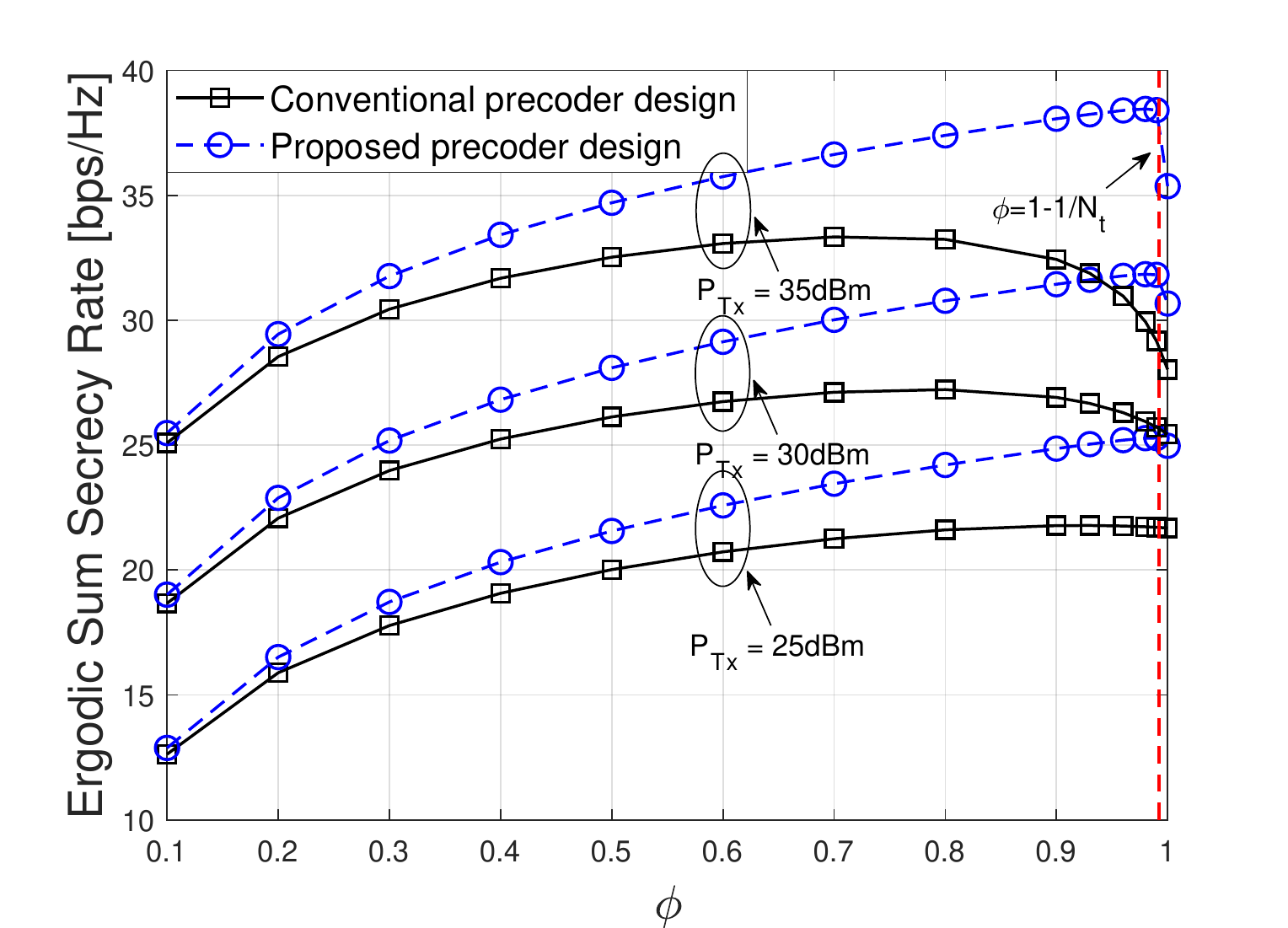}\label{fig:SR_Ptx_Nt128}}
	\caption{Ergodic sum secrecy rate depending on the transmit power when $\kappa=10$~dB, $\sigma_{\epsilon}=0^{\circ}$. } 
	\label{fig:SR_Ptx}
\end{figure}
\begin{figure}[t]
	\centering
	\subfloat[$N_{\mathsf{t}}=16$]{
\includegraphics[width=0.5\textwidth]{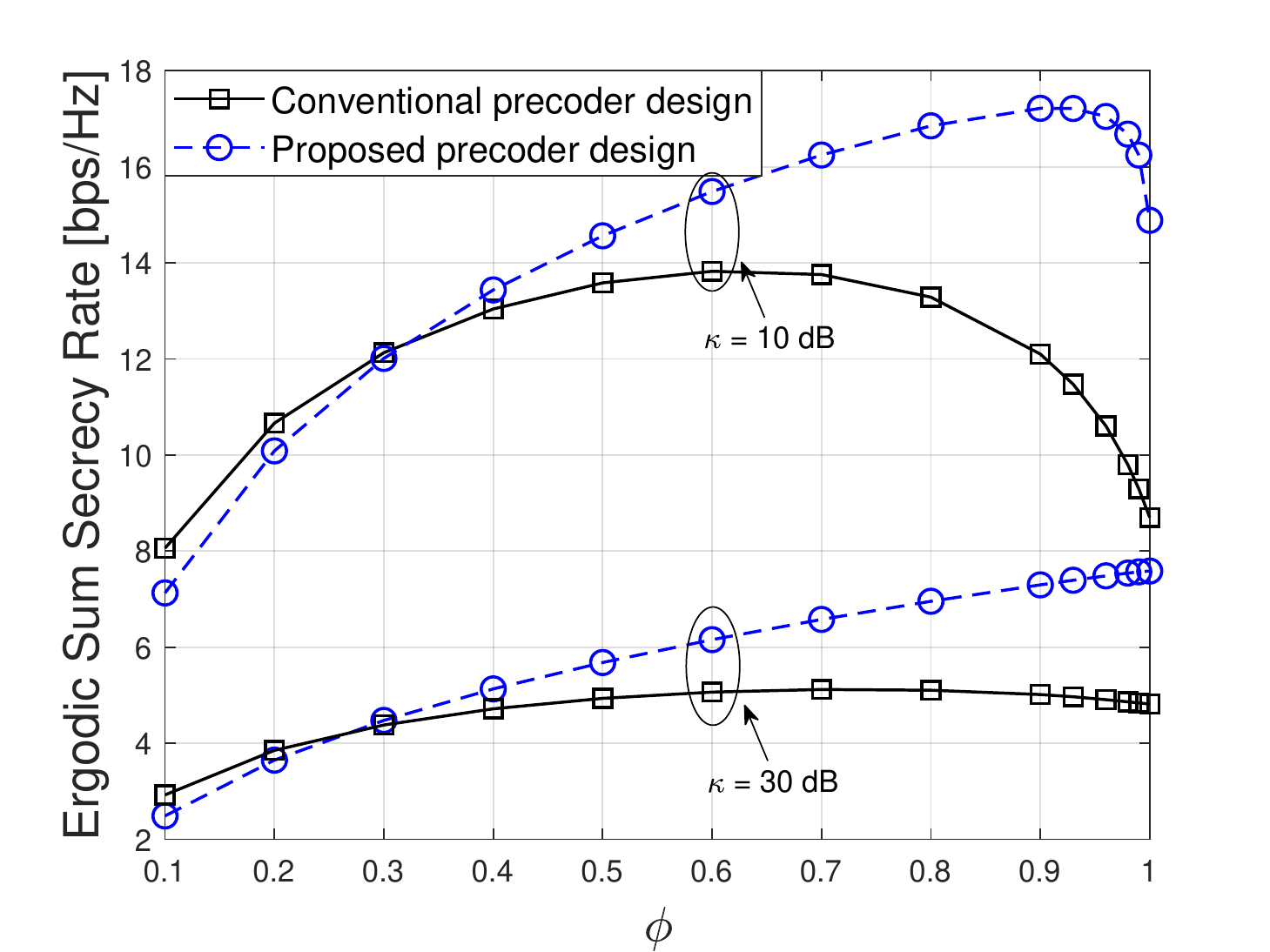}\label{fig:SR_KF_Nt16}}
\vspace{-0.01in}
	\subfloat[$N_{\mathsf{t}}=128$]{
	\includegraphics[width=0.5\textwidth]{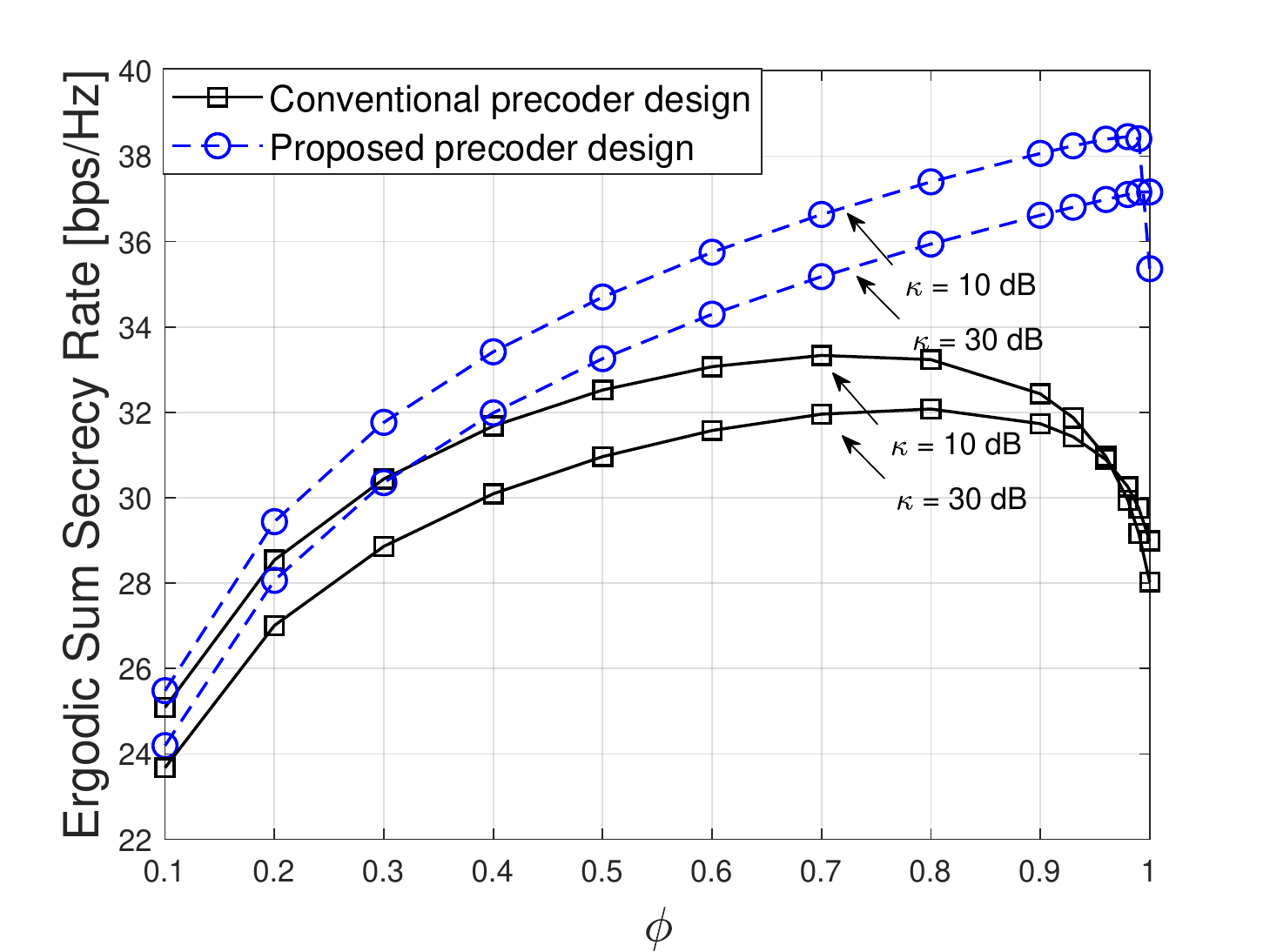}\label{fig:SR_KF_Nt128}}
	\caption{Ergodic sum secrecy rate depending on the Rician K-factor when $\mathsf{P}_{\mathsf{Tx}}=35$~dB, $\sigma_{\epsilon}=0^{\circ}$. } 
	\label{fig:SR_KF}
\end{figure}
\begin{figure}[t]
	\centering
	\subfloat[$N_{\mathsf{t}}=16$]{
\includegraphics[width=0.5\textwidth]{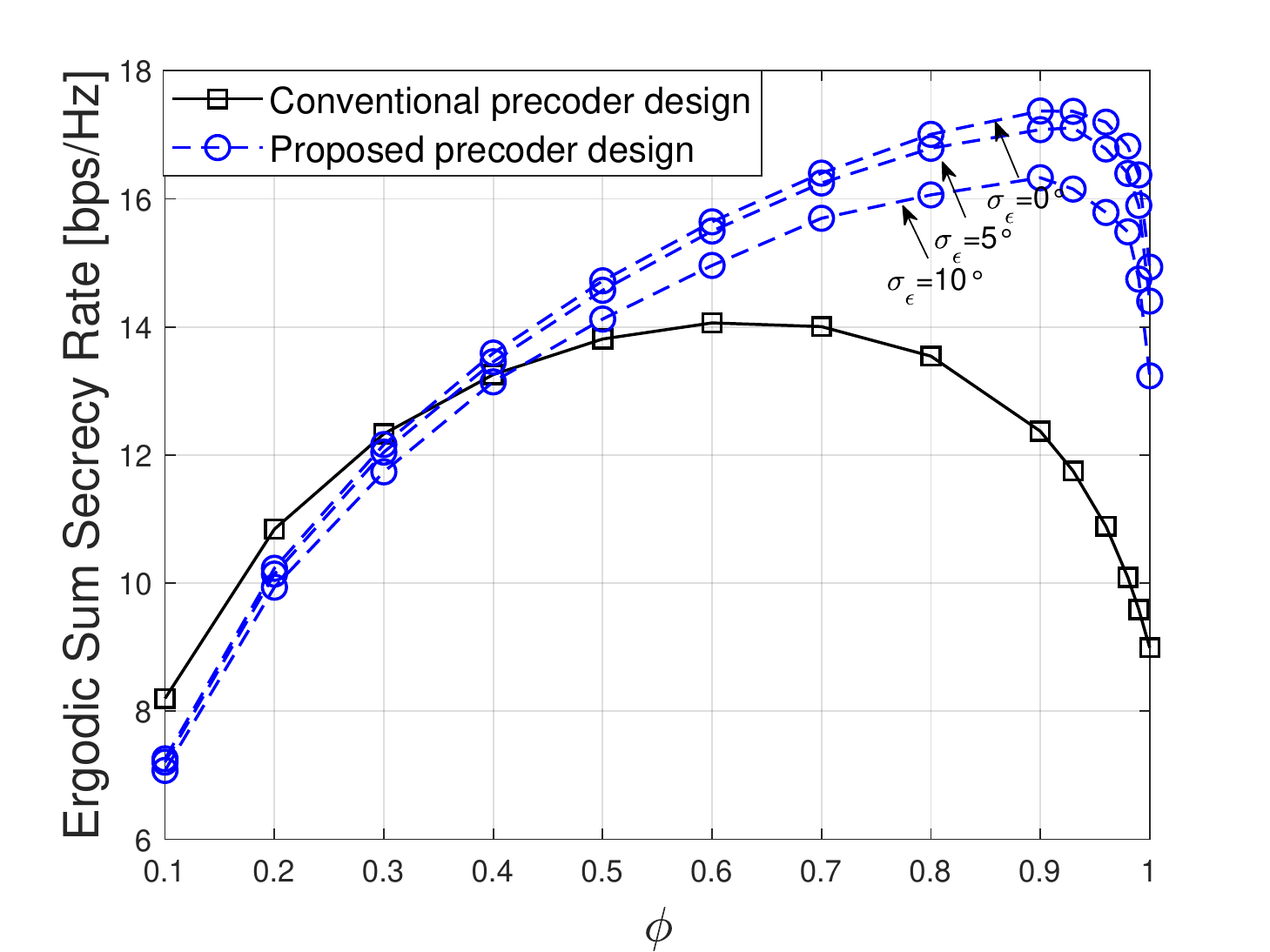}\label{fig:SR_eps_Nt16}}
\vspace{-0.01in}
	\subfloat[$N_{\mathsf{t}}=128$]{
	\includegraphics[width=0.5\textwidth]{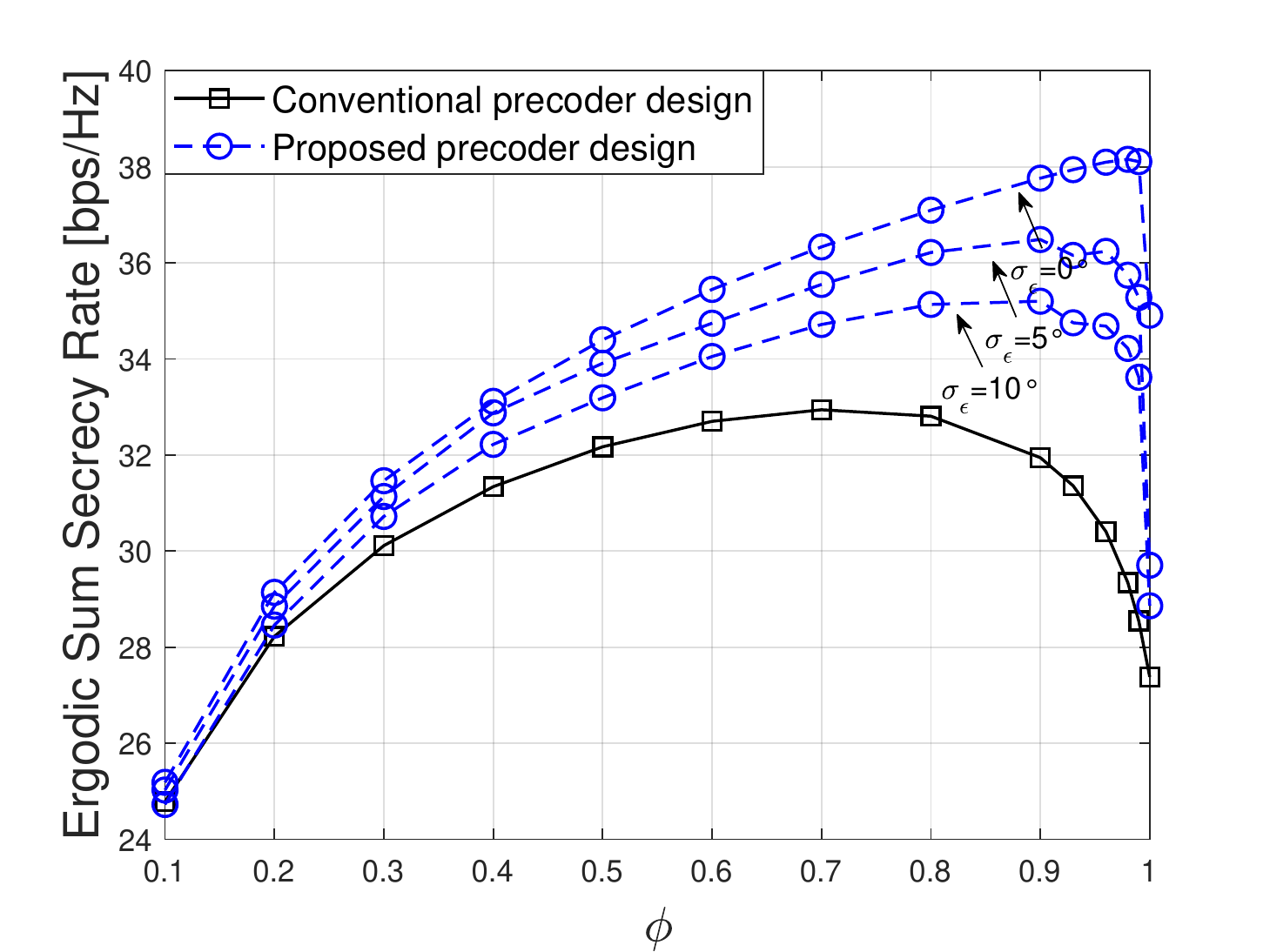}\label{fig:SR_eps_Nt128}}
	\caption{Ergodic sum secrecy rate depending on the standard variance of angular calibration error ($\sigma_{\epsilon}$) when $\mathsf{P}_{\mathsf{Tx}}=35$~dB, $\kappa=10$~dB. } 
	\label{fig:SR_eps}
\end{figure}
\begin{figure}[t]
	\centering
\includegraphics[width=0.5\textwidth]{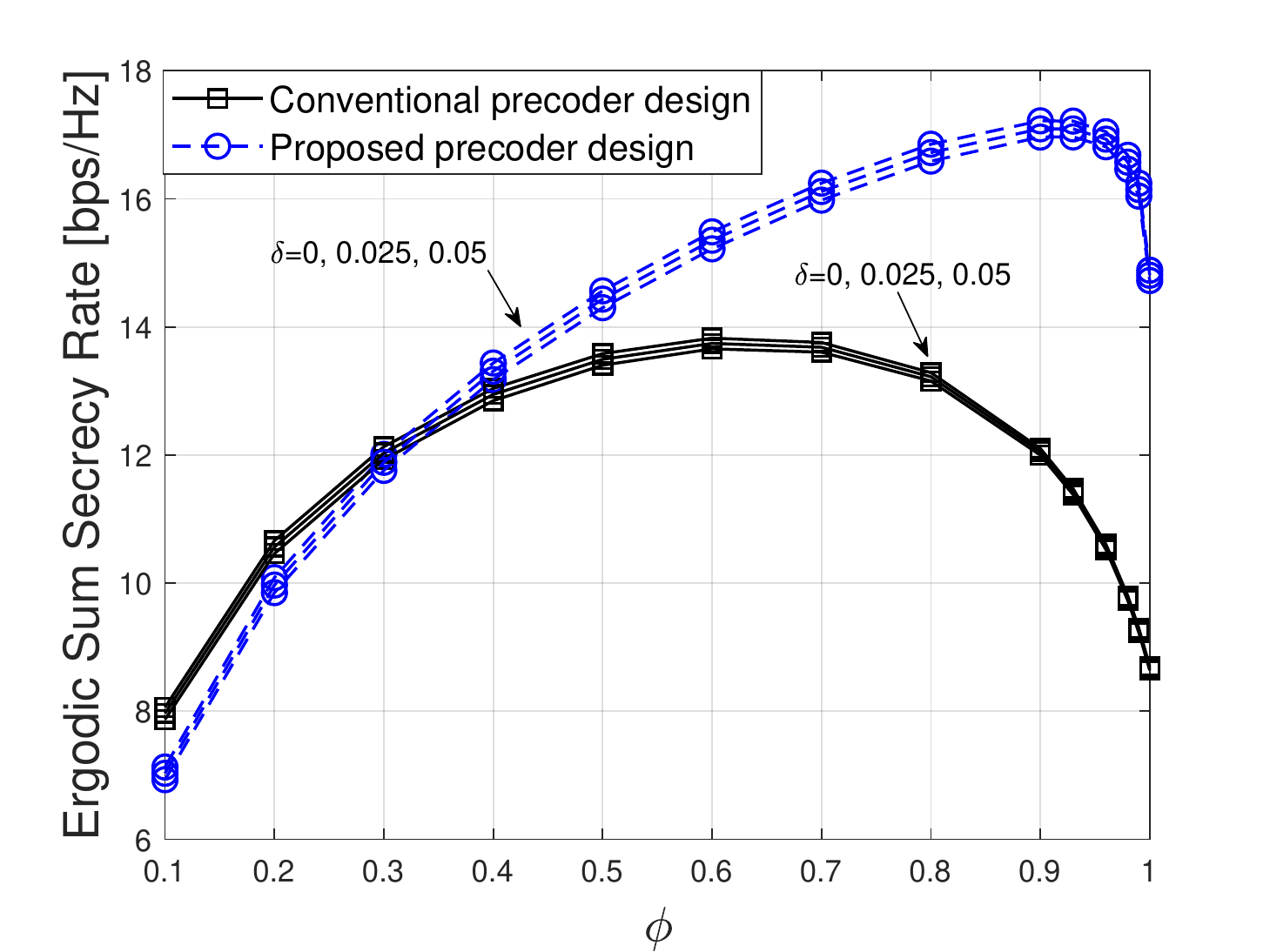}
	\caption{Ergodic sum secrecy rate depending on the tag power factor ($\delta$) when $\mathsf{P}_{\mathsf{Tx}}=35$~dB, $N_{\mathsf{t}}=16$, $\kappa=10$~dB, $\sigma_{\epsilon}=0^{\circ}$. } 
	\label{fig:SR_delta}
\end{figure}

Fig.~\ref{fig:SR_Ptx} shows the ergodic sum secrecy rate depending on the transmit power ($\mathsf{P}_{\mathsf{Tx}}$), where the transmit power is controlled by changing $\phi$ as described in \eqref{eq:received_signal:user}. We consider the results with the moderate number of antennas ($N_{\mathsf{t}}=16$) and the large number of antennas ($N_{\mathsf{t}}=128$) in Fig.~\ref{fig:SR_Ptx_Nt16} and Fig.~\ref{fig:SR_Ptx_Nt128}, respectively. We first observe that the performance of the proposed precoder design is superior to the conventional precoder design regardless of the transmit power, which indicates that our proposed precoder design is robust to the transmit power. Secondly, for the moderate number of the antennas case, we can observe that the conventional precoder design is slightly better than the proposed one in the low $\phi$ range, but with large number of antennas, the proposed precoder always outperforms the conventional precoder for the whole $\phi$ range. We can explain it by \textit{Corollary~1} and \textit{Corollary~2}: the achievable rate of the UAV-UE for the conventional precoder design is greater than that of the proposed precoder design but the performance gap disappears as $N_{\mathsf{t}}$ grows without bound. In the end, the ergodic secrecy rate for the proposed precoder design is always better than the conventional precoder design by \textit{Corollary~4}. Thirdly, it is observed that the optimal point that maximizes the ergodic sum secrecy rate is  $\phi=1-\frac{1}{N_{\mathsf{t}}}$ in both the moderate and large number of antennas, which means that the asymptotic analysis in \textit{Corollary~6} holds quite well even the moderate number of antennas. Note that the curve of the ergodic sum secrecy rate can be an increasing function of $\phi$, if the ergodic achievable rate of the UAV-Eve is too small to make it concave, which is shown for the case of $\mathsf{P}_{\mathsf{Tx}}=25$~dB with  moderate number of antennas.

\subsection{Ergodic Sum Secrecy Rate Depending on Rician K-factor}

In Fig.~\ref{fig:SR_KF}, we show the ergodic sum secrecy rate depending on the K-factor ($\kappa$). We consider K-factor = $10$~dB and $30$~dB as the moderate and the high K-factor values, respectively. It is observed that the proposed precoder design achieves better performance compared with the conventional one for both K-factors. In addition, we also observe that when $\kappa=30$~dB, the ergodic sum secrecy rate keeps increasing until $\phi$ reaches $1$ in the proposed precoder design. This is due to \textit{Corollary~5} that the achievable rate of the UAV-Eve asymptotically goes to zero as $\kappa$ grows without bound and the ergodic sum secrecy rate becomes an increasing function of $\phi$.

\subsection{The Effect of Angular Calibration Error and Tag Power Factor on the Ergodic Sum Secrecy Rate}

Fig.~\ref{fig:SR_eps} shows the effect of angular calibration error ($\epsilon$) on the ergodic sum secrecy rate. The performance degradation is clearly observed as the variance of the error increases. Moreover, the degradation is bigger for a large number of antennas when compared with the case for moderate number of antennas. These observations are expected from \eqref{eq:MSE_ang_cal}, which explains the dependency of these parameters on the amount of the MSE.

In Fig.~\ref{fig:SR_delta}, we show the effect of a tag power factor on the ergodic secrecy rate. Although we optimize the tag power factor ($\delta$) in \eqref{eq:optimization_ssr_1}, we need to ensure that the amount of $\delta$ that we obtain does not degrade the secrecy rate much. It is observed that the small amount of the power on the authentication tag by $\delta$ slightly degrades the performance. However, it shows that the effect on ergodic sum secrecy rate is trivial even at $\delta=0.05$.

\subsection{Secret Key Recovery Probability of UAV-Eve}

\begin{figure}[t]
	\centering
\includegraphics[width=0.5\textwidth]{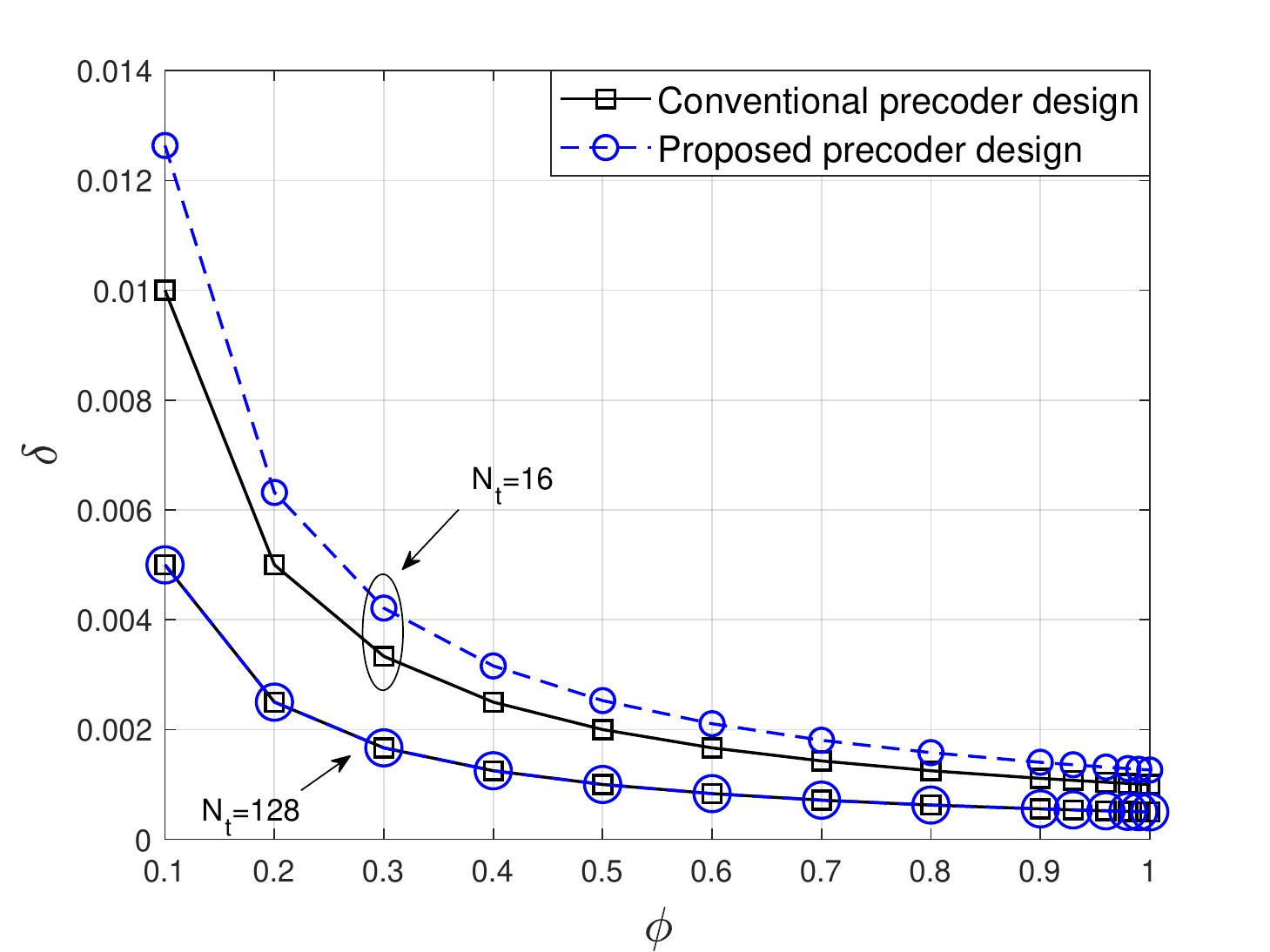}
	\caption{The optimized tag power factor ($\delta$) by Algorithm 1 when $\mathsf{P}_{\mathsf{Tx}}=35$~dB, $\kappa=10$~dB, $\sigma_{\epsilon}=0^{\circ}$. } 
	\vspace{-0.1in}
	\label{fig:delta_opt}
\end{figure}
\begin{figure}[t]
	\centering
\includegraphics[width=0.5\textwidth]{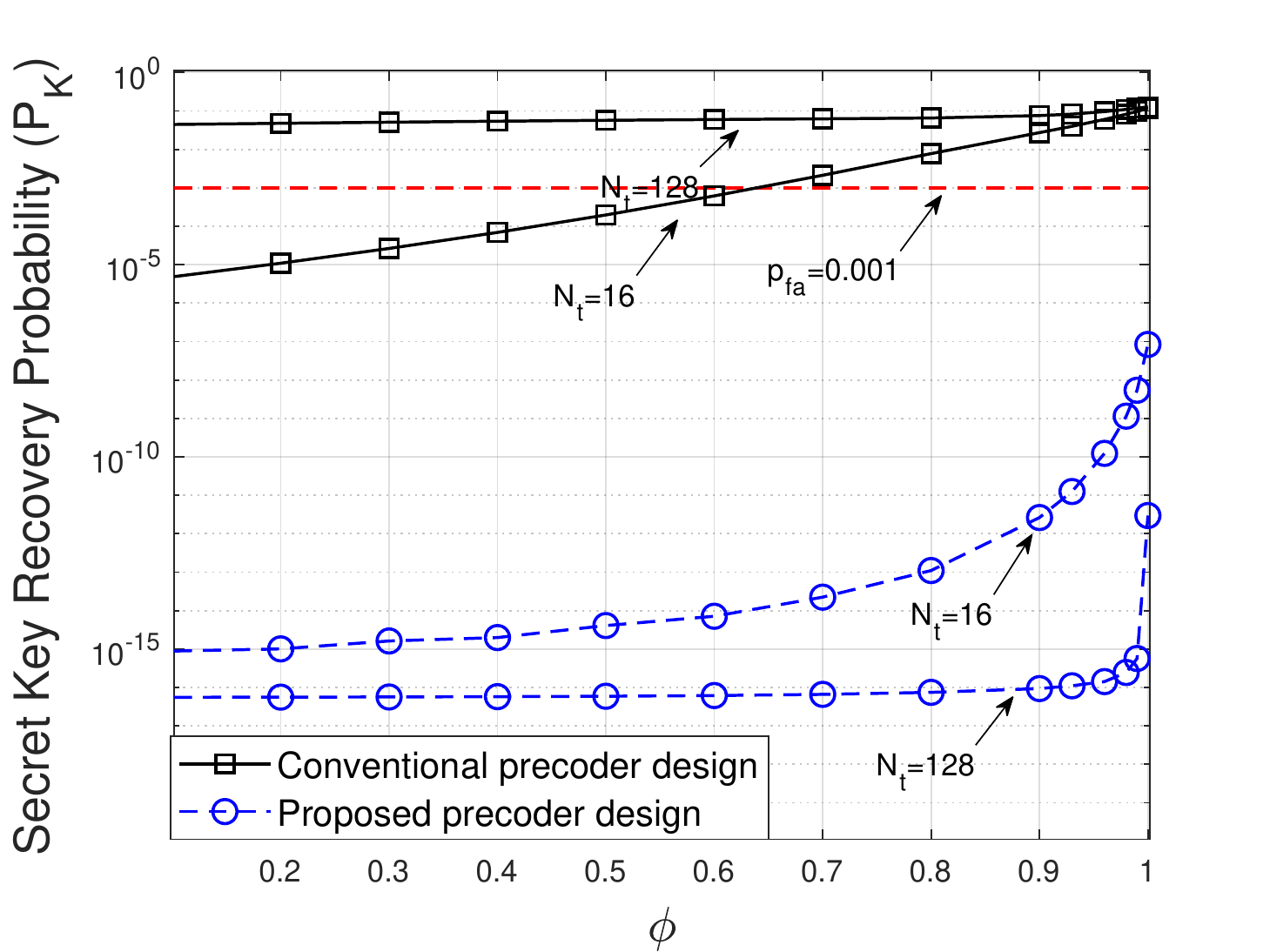}
	\caption{The secret key recovery probability  when $\mathsf{P}_{\mathsf{Tx}}=35$~dB, $\kappa=10$~dB, $\sigma_{\epsilon}=0^{\circ}$. } 
	\label{fig:PK}
\end{figure}

Fig.~\ref{fig:delta_opt} shows the optimal tag power ($\delta_{\mathsf{opt}}$) that is obtained using Algorithm 1. It is observed that the optimized value of $\delta$ for both the moderate and large number of antennas cases are below 0.015. Since we already observe from Fig.~\ref{fig:SR_delta} that $\delta=0.025$ is not enough to degrade secrecy rate, $\delta_{\mathsf{opt}}$ will not degrade the ergodic sum secrecy rate as well. We also observe that the optimal $\delta$ is the rational function of $\phi$, which can be expected from line 7 of Algorithm 1.

In Fig.~\ref{fig:PK}, we show the secret key recovery probability of the UAV-Eve in \eqref{eq:P_K}. We compare the performance with the false alarm probability ($p_{\mathsf{fa}}$) of the authentication probability in \eqref{eq:P_A} which is the same as the probability of success of the random tag attack from the UAV-Eve. If $P_{\mathsf{K}}$ is higher than $p_{\mathsf{fa}}$, we can no longer guarantee the security level that the random tag attack can achieve. Note that UAV-Eve is able to randomly generate a tag from the keyspace and transmit it to a user to impersonate GS. The probability that the secret key is recovered at the UAV-Eve for the conventional precoder design is fairly higher than that for the proposed precoder design in both the moderate and the large number of antennas cases. Moreover, the probability for the conventional precoder design is higher than the false alarm probability, which means that the probability that UAV-Eve guesses the correct secret key is higher than UAV-Eve can guess it by a random tag attack. On the other hand, the proposed precoder design protects well from the attack from the UAV-Eve.

\section{Conclusion}\label{sec:conclusion}
In this paper, we study the precoder designs for the physical-layer security and the authentication on UAV massive MIMO networks. We propose a precoder design considering the imperfect location information of the UAV-Eve and compare the performance with a conventional precoder design. We obtain the closed-form expressions of the ergodic achievable rate as well as a large number of antennas limit for both the conventional and the proposed precoder designs and analytically show the superiority of the proposed precoder design. Furthermore, we optimize the power splitting factor by the asymptotical analysis. We also adopt the fingerprint embedding authentication framework and optimize the tag power factor. By simulation results, we show that the proposed precoder design outperforms the conventional precoder design in terms of the ergodic sum secrecy rate as well as the physical layer authentication framework.
 
\appendices

\section{Proof of Theorem 1}

We first derive $\mathbb|\textbf{h}_u^{\rm H}\textbf{w}_u^{\mathsf{cv}}|^2$ as follows:
\begin{align}\label{eq:theo1:p1}
    |\textbf{h}_u^{\rm H}\textbf{w}_u^{\mathsf{cv}}|^2&\overset{(a)}{=}\frac{1}{K\|\Tilde{\textbf{w}}_u^{\mathsf{cv}}\|^2} 
    =\frac{1}{K\left[\left(\Tilde{\textbf{W}}^{\mathsf{cv}}\right)^{\rm H}\Tilde{\textbf{W}}^{\mathsf{cv}}\right]_{uu}}\nonumber\\
    &\overset{(b)}{=}\frac{1}{K[(\textbf{H}^{\rm H}\textbf{H})^{-1}]_{uu}},
\end{align}
where $(a)$ comes from \eqref{eq:ZF_precoder_norm:conv} and $\textbf{h}_u^{\rm H}\Tilde{\textbf{w}}_u^{\mathsf{cv}}=1$, $(b)$ comes from \eqref{eq:ZF_precoder:conv}, and $[\textbf{X}]_{uu}$ indicates $u_{\rm th}$ diagonal entry of matrix $\textbf{X}$. Note that $\textbf{H}^{\rm H}\textbf{H}$ follows a non-central Wishart distribution and it is not tractable to derive analytical expression. However, we can approximate it to a central Wishart $K\times K$  complex-valued random matrix with $N_{\mathsf{t}}$ degrees of freedom with covariance matrix $\boldsymbol{\Sigma}_{\textbf{H}}$ as \cite[Theroem 4]{zhang2014power}, \cite{siriteanu2012mimo}
\begin{align}
    \textbf{H}^{\rm H}\textbf{H}&\sim\mathcal{W}_K(N_{\mathsf{t}},\boldsymbol{\Sigma}_{\textbf{H}}),
\end{align}
where the covariance matrix $\boldsymbol{\Sigma}_{\textbf{H}}$ is given by 
\begin{align}\label{eq:theo1:p4}
 \boldsymbol{\Sigma}_{\textbf{H}}&=\frac{1}{\kappa+1}\textbf{I}+\frac{\kappa}{N_{\mathsf{t}}(\kappa+1)}\textbf{H}_{\mathsf{LoS}}^{\rm H}\textbf{H}_{\mathsf{LoS}},
\end{align}
where $\textbf{H}_{\mathsf{LoS}}=\left[\textbf{h}_{\mathsf{LoS},1},\dots,\textbf{h}_{\mathsf{LoS},K}\right]$ is the aggregate matrix of the LoS component of the UAV-UEs channel. Then, under the above approximation, $1/[(\textbf{H}^{\rm H}\textbf{H})^{-1}]_{uu}$ follows a Chi-squared distribution and we can obtain
\begin{align}
    \mathbb{E}\left\{[(\mathbf{H}^{\rm H}\mathbf{H})^{-1}]_{uu}\right\}&=\frac{[\boldsymbol{\Sigma}^{-1}_{\textbf{H}}]_{uu}}{N_{\mathsf{t}}-K}.
\end{align}
By plugging it in \eqref{eq:rate_user}, \eqref{eq:theo1:p1}, we can obtain \eqref{eq:theo1:def}.

\section{Proof of Theorem 2}

Similar to \textit{Theorem 1}, we first derive $\mathbb|\textbf{h}_u^{\rm H}\textbf{w}_u^{\mathsf{pp}}|^2$ as follows:
\begin{align}\label{eq:theo2:p1}
    |\textbf{h}_u^{\rm H}\textbf{w}_u^{\mathsf{pp}}|^2&=\frac{1}{K\|\Tilde{\textbf{w}}_u^{\mathsf{pp}}\|^2}
    =\frac{1}{K\left[\left(\Tilde{\textbf{W}}^{\mathsf{pp}}\right)^{\rm H}\Tilde{\textbf{W}}^{\mathsf{pp}}\right]_{uu}}\nonumber\\
    &=\frac{1}{K[(\textbf{G}^{\rm H}\textbf{G})^{-1}]_{uu}}.
\end{align}
By using the way in \textit{Theorem 1}, a non-central Wishart complex-valued matrix $\textbf{G}^{\rm H}\textbf{G}$ can be approximated as a central Wishart distribution $(K+1)\times (K+1)$ complex-valued matrix as follows:  
\begin{align}
    \textbf{G}^{\rm H}\textbf{G}&\sim\mathcal{W}_{K+1}(N_{\mathsf{t}},\boldsymbol{\Sigma}_{\textbf{G}}),
\end{align}
where the covariance matrix $\boldsymbol{\Sigma}_{\textbf{G}}$ can be expressed from \eqref{eq:theo1:p4} as the following block-wise matrix
\begin{align}
 \boldsymbol{\Sigma}_{\textbf{G}}&=\left[ \begin{array}{c|c}
        \boldsymbol{\Sigma}_{\textbf{H}} & \frac{\sqrt{\kappa}}{N_{\mathsf{t}}\sqrt{\kappa+1}}\textbf{H}_{\mathsf{LoS}}^{\rm H}\textbf{g}_{\rm e} \\
        \hline
        \frac{\sqrt{\kappa}}{N_{\mathsf{t}}\sqrt{\kappa+1}}\textbf{g}_{\rm e}^{\rm H}\textbf{H}_{\mathsf{LoS}} & \frac{1}{N}\textbf{g}_{\rm e}^{\rm H}\textbf{g}_{\rm e}
    \end{array}\right]\nonumber\\
    &=\left[ \begin{array}{c|c}
        \boldsymbol{\Sigma}_{\textbf{H}} & \frac{\sqrt{\kappa}}{N_{\mathsf{t}}\sqrt{\kappa+1}}\textbf{H}_{\mathsf{LoS}}^{\rm H}\textbf{g}_{\rm e} \\
        \hline
        \frac{\sqrt{\kappa}}{N_{\mathsf{t}}\sqrt{\kappa+1}}\textbf{g}_{\rm e}^{\rm H}\textbf{H}_{\mathsf{LoS}} & 1
    \end{array}\right]\label{eq:theo2:p3}\\
    &=\left[ \begin{array}{c|c}
        \boldsymbol{\Sigma}_{\textbf{H}} & \textbf{b} \\
        \hline
        \textbf{c} & 1
    \end{array}\right].\label{Eq_32}
\end{align}
The expression above follows from $\textbf{G}=[\textbf{H}\; \textbf{g}_{\rm e}]$, and $\textbf{b}$, $\textbf{c}$ are the substitution of the block-wise expression in \eqref{eq:theo2:p3}. By using the block-wise inversion formula \cite[Proposition 2.8.7]{bernstein2009matrix}, we can obtain the following expression,
\begin{align}\label{eq:theo2:p5}
    [\boldsymbol{\Sigma}_{\textbf{G}}^{-1}]_{uu}&=[\boldsymbol{\Sigma}_{\textbf{H}}^{-1}]_{uu}+[\boldsymbol{\Sigma}_{\textbf{H}}^{-1}\textbf{b}\textbf{c}\boldsymbol{\Sigma}_{\textbf{H}}^{-1}(1-\textbf{c}\boldsymbol{\Sigma}_{\textbf{H}}^{-1}\textbf{b})^{-1}]_{uu}\nonumber\\
    &=[\boldsymbol{\Sigma}_{\textbf{H}}^{-1}]_{uu}+\Delta,
\end{align}
where the second term is replaced with $\Delta$. Note that the index $u=1,\dots,K$, and the last $(K+1)_{\rm th}$ diagonal entry of $\boldsymbol{\Sigma}_{\textbf{G}}^{-1}$ is given differently from the block-wise inversion formula as
\begin{align}\label{eq:theo2:p6}
    [\boldsymbol{\Sigma}_{\textbf{G}}^{-1}]_{(K+1)(K+1)}&=[\boldsymbol{\Sigma}_{\textbf{G}}^{-1}]_{\mathsf{ee}}=(1-\textbf{c}\boldsymbol{\Sigma}_{\textbf{H}}^{-1}\textbf{b})^{-1}.
\end{align}
Note that we replace the index $(K+1)$ to `$\mathsf{e}
$', since the last diagonal entry contributes to the UAV-Eve direction. By the same approach with \textit{Theorem 1}, $1/[(\textbf{G}^{\rm H}\textbf{G})^{-1}]_{uu}$ follows a Chi-squared distribution and we can obtain from \eqref{eq:theo2:p5} 
\begin{align}\label{eq:theo2:p7}
    \mathbb{E}\left\{[(\mathbf{G}^{\rm H}\mathbf{G})^{-1}]_{uu}\right\}&=\frac{[\boldsymbol{\Sigma}^{-1}_{\textbf{G}}]_{uu}}{N_{\mathsf{t}}-K-1}\nonumber\\
    &=\frac{[\boldsymbol{\Sigma}^{-1}_{\textbf{H}}]_{uu}+\Delta}{N_{\mathsf{t}}-K-1}.
\end{align}
By plugging it in \eqref{eq:rate_user}, \eqref{eq:theo2:p1}, we can obtain \eqref{eq:theo2:def}.

\section{Proof of Theorem 3}

We first rewrite $\mathbb{E}\left\{|\textbf{h}_\mathsf{e}^{\rm H}\textbf{w}_u^{\mathsf{cv}}|^2\right\}$ as follows:
\begin{align}\label{eq:theo3:p1}
    &\mathbb{E}\{|\textbf{h}_\mathsf{e}^{\rm H}\textbf{w}_u^{\mathsf{cv}}|^2\}\nonumber\\
    &=\mathbb{E}\left\{\left|\left(\sqrt{ \frac{ \kappa }{ \kappa + 1 } }\textbf{h}_{\mathsf{LoS},\rm e}+\sqrt{ \frac{ 1 }{ \kappa + 1 }}\textbf{h}_{\mathsf{NLoS},\rm e}\right)^{\rm H}\textbf{w}_u^{\mathsf{cv}}\right|^2\right\},
\end{align}
where the expression comes from \eqref{eq:Rician_eve}. Then, we also get
\begin{align}
    &\mathbb{E}\{|\textbf{h}_\mathsf{e}^{\rm H}\textbf{w}_u^{\mathsf{cv}}|^2\}\nonumber\\
    &=\mathbb{E}\left\{\frac{ \kappa }{ \kappa + 1 }\left|\textbf{h}_{\mathsf{LoS},\rm e}^{\rm H}\textbf{w}_u^{\mathsf{cv}}\right|^2+\frac{ \sqrt{\kappa} }{ \kappa + 1 }\left(\textbf{h}_{\mathsf{LoS},\rm e}^{\rm H}\textbf{w}_u^{\mathsf{cv}}\right)\left(\textbf{h}_{\mathsf{NLoS},\rm e}^{\rm H}\textbf{w}_u^{\mathsf{cv}}\right)^{\ast}\nonumber\right.\\
    &\left.+\frac{ \sqrt{\kappa} }{ \kappa + 1 }\left(\textbf{h}_{\mathsf{LoS},\rm e}^{\rm H}\textbf{w}_u^{\mathsf{cv}}\right)^{\ast}\left(\textbf{h}_{\mathsf{NLoS},\rm e}^{\rm H}\textbf{w}_u^{\mathsf{cv}}\right)+
    \frac{ 1 }{ \kappa + 1 }\left|\textbf{h}_{\mathsf{NLoS},\rm e}^{\rm H}\textbf{w}_u^{\mathsf{cv}}\right|^2\right\}.
\end{align}
By removing the terms with zero mean, we can rewrite it as
\begin{align}\label{eq:theo3:p3}
    &\mathbb{E}\{|\textbf{h}_\mathsf{e}^{\rm H}\textbf{w}_u^{\mathsf{cv}}|^2\}\nonumber\\
    &=\frac{ \kappa }{ \kappa + 1 }\mathbb{E}\left\{\left|\textbf{h}_{\mathsf{LoS},\rm e}^{\rm H}\textbf{w}_u^{\mathsf{cv}}\right|^2\right\}+\frac{ 1 }{ \kappa + 1 }\mathbb{E}\left\{\left|\textbf{h}_{\mathsf{NLoS},\rm e}^{\rm H}\textbf{w}_u^{\mathsf{cv}}\right|^2\right\}.
\end{align}
The first term of \eqref{eq:theo3:p3} can be simplify as
\begin{align}\label{eq:theo3:p4}
&\frac{ \kappa }{ \kappa + 1 }\mathbb{E}\left\{\left|\textbf{h}_{\mathsf{LoS},\rm e}^{\rm H}\textbf{w}_u^{\mathsf{cv}}\right|^2\right\}=\frac{ \kappa }{ \kappa + 1 }\mathbb{E}\left\{\left(\textbf{w}_u^{\mathsf{cv}}\right)^{\rm H}\textbf{h}_{\mathsf{LoS},\rm e}\textbf{h}_{\mathsf{LoS},\rm e}^{\rm H}\textbf{w}_u^{\mathsf{cv}}\right\}\nonumber\\
&=\frac{ \kappa }{ \kappa + 1 }\text{Tr}\left[\textbf{h}_{\mathsf{LoS},\rm e}\textbf{h}_{\mathsf{LoS},\rm e}^{\rm H}\mathbb{E}\left\{\textbf{w}_u^{\mathsf{cv}}\left(\textbf{w}_u^{\mathsf{cv}}\right)^{\rm H})\right\}\right]\nonumber\\
&=\frac{ \kappa }{ \kappa + 1 }\text{Tr}\left[\textbf{h}_{\mathsf{LoS},\rm e}\textbf{h}_{\mathsf{LoS},\rm e}^{\rm H}\Gamma_{\textbf{w}}\right],
\end{align}
where $\Gamma_{\textbf{w}}=\mathbb{E}\left\{\textbf{w}_u^{\mathsf{cv}}\left(\textbf{w}_u^{\mathsf{cv}}\right)^{\rm H})\right\}$. Then, the second term of \eqref{eq:theo3:p3} can be rewritten as
\begin{align}\label{eq:theo3:p5}
    \frac{ 1 }{ \kappa + 1 }\mathbb{E}\left\{\left|\textbf{h}_{\mathsf{NLoS},\rm e}^{\rm H}\textbf{w}_u^{\mathsf{cv}}\right|^2\right\}\overset{(a)}{=}\frac{ 1 }{ \kappa + 1 }\|\textbf{w}_u^{\mathsf{cv}}\|^2=\frac{ 1 }{ K( \kappa + 1) },
\end{align}
where (a) comes from the fact that $\textbf{w}_u^{\mathsf{cv}}$ is independent of the small-scale fading vector $\textbf{h}_{\mathsf{NLoS},\rm e}$ and $\left|\textbf{h}_{\mathsf{NLoS},\rm e}^{\rm H}\textbf{w}_u^{\mathsf{cv}}\right|^2$ follows (scaled) chi-square distribution with 2 degree of freedom \cite[Lemma 2]{zhu2014secure}.

Next, we calculate $\sum_{i=1}^{N_{\mathsf{AN}}}\mathbb{E}\left\{|\textbf{h}_\mathsf{e}^{\rm H}\textbf{v}_i^{\mathsf{cv}}|^2\right\}$ as follows:
\begin{align}
    &\sum_{i=1}^{N_{\mathsf{AN}}}\mathbb{E}\left\{|\textbf{h}_\mathsf{e}^{\rm H}\textbf{v}_i^{\mathsf{cv}}|^2\right\}\nonumber\\
    &=\sum_{i=1}^{N_{\mathsf{AN}}}\mathbb{E}\left\{\left|\left(\sqrt{ \frac{ \kappa }{ \kappa + 1 } }\textbf{h}_{\mathsf{LoS},\rm e}+\sqrt{ \frac{ 1 }{ \kappa + 1 }}\textbf{h}_{\mathsf{NLoS},\rm e}\right)^{\rm H}\textbf{v}_i^{\mathsf{cv}}\right|^2\right\}.
\end{align}
By using similar calculation with $\mathbb{E}\{|\textbf{h}_\mathsf{e}^{\rm H}\textbf{w}_u^{\mathsf{cv}}|^2\}$, we can get
\begin{align}\label{eq:theo3:p7}
    &\sum_{i=1}^{N_{\mathsf{AN}}}\mathbb{E}\left\{|\textbf{h}_\mathsf{e}^{\rm H}\textbf{v}_i^{\mathsf{cv}}|^2\right\}\nonumber\\
    &=\frac{ \kappa }{ \kappa + 1 }\mathbb{E}\left\{\sum_{i=1}^{N_{\mathsf{AN}}}\left|\textbf{h}_{\mathsf{LoS},\rm e}^{\rm H}\textbf{v}_i^{\mathsf{cv}}\right|^2\right\}+\frac{ 1 }{ \kappa + 1 }\mathbb{E}\left\{\sum_{i=1}^{N_{\mathsf{AN}}}\left|\textbf{h}_{\mathsf{NLoS},\rm e}^{\rm H}\textbf{v}_i^{\mathsf{cv}}\right|^2\right\}\nonumber\\
    &=\frac{ \kappa }{ \kappa + 1 }\mathbb{E}\left\{\textbf{h}_{\mathsf{LoS},\rm e}^{\rm H}\textbf{V}^{\mathsf{cv}}\left(\textbf{V}^{\mathsf{cv}}\right)^{\rm H}\textbf{h}_{\mathsf{LoS},\rm e}\right\}+\frac{ 1 }{ \kappa + 1 }\sum_{i=1}^{N_{\mathsf{AN}}}\left\|\textbf{v}_i^{\mathsf{cv}}\right\|^2\nonumber\\
    &=\frac{ \kappa }{ \kappa + 1 }\text{Tr}\left[\textbf{h}_{\mathsf{LoS},\rm e}\textbf{h}_{\mathsf{LoS},\rm e}^{\rm H}\Gamma_{\textbf{V}}\right]+\frac{ 1 }{ \kappa + 1 },
\end{align}
where $\Gamma_{\textbf{V}}=\mathbb{E}\left\{\textbf{V}^{\mathsf{cv}}\left(\textbf{V}^{\mathsf{cv}}\right)^{\rm H}\right\}$. Finally, by plugging \eqref{eq:theo3:p3}, \eqref{eq:theo3:p4}, \eqref{eq:theo3:p5}, \eqref{eq:theo3:p7} in \eqref{eq:rate_eve}, we can obtain \eqref{eq:theo3:def}.

\section{Proof of Theorem 4}

We first calculate $\mathbb{E}\left\{|\textbf{h}_\mathsf{e}^{\rm H}\textbf{w}_u^{\mathsf{pp}}|^2\right\}$ as follow
\begin{align}\label{eq:theo4:p1}
    &\mathbb{E}\{|\textbf{h}_\mathsf{e}^{\rm H}\textbf{w}_u^{\mathsf{pp}}|^2\}\nonumber\\
    &=\mathbb{E}\left\{\left|\left(\sqrt{ \frac{ \kappa }{ \kappa + 1 } }\textbf{h}_{\mathsf{LoS},\rm e}+\sqrt{ \frac{ 1 }{ \kappa + 1 }}\textbf{h}_{\mathsf{NLoS},\rm e}\right)^{\rm H}\textbf{w}_u^{\mathsf{pp}}\right|^2\right\}\nonumber\\
    &\overset{(a)}{=}\frac{ 1 }{ \kappa + 1 }\mathbb{E}\left\{\left|\left(\textbf{h}_{\mathsf{NLoS},\rm e}\right)^{\rm H}\textbf{w}_u^{\mathsf{pp}}\right|^2\right\}
\end{align}
where (a) comes from the fact that $\left(\textbf{h}_{\mathsf{LoS},\rm e}\right)^{\rm H}\textbf{w}_u^{\mathsf{pp}}=0$. Note that the $\textbf{w}_u^{\mathsf{pp}}$ is designed by the ZF precoding which nullifies $\textbf{h}_{\mathsf{LoS},\rm e}$ vector by treating it as  other channel vector. By using \cite[Lemma 2]{zhu2014secure}, we can also obtain 
\begin{align}\label{eq:theo4:p2}
    \mathbb{E}\{|\textbf{h}_\mathsf{e}^{\rm H}\textbf{w}_u^{\mathsf{pp}}|^2\}=\frac{1}{\kappa+1}\|\textbf{w}_u^{\mathsf{pp}}\|^2=\frac{1}{K(\kappa+1)}.
\end{align}

Next, we derive $\sum_{i=1}^{N_{\mathsf{AN}}}\mathbb{E}\{|\textbf{h}_\mathsf{e}^{\rm H}\textbf{v}_i^{\mathsf{pp}}|^2\}$ as follow
\begin{align}
    &\sum_{i=1}^{N_{\mathsf{AN}}}\mathbb{E}\left\{|\textbf{h}_\mathsf{e}^{\rm H}\textbf{v}_i^{\mathsf{pp}}|^2\right\}\nonumber\\
    &=\mathbb{E}\left\{\left|\left(\sqrt{ \frac{ \kappa }{ \kappa + 1 } }\textbf{h}_{\mathsf{LoS},\rm e}+\sqrt{ \frac{ 1 }{ \kappa + 1 }}\textbf{h}_{\mathsf{NLoS},\rm e}\right)^{\rm H}\textbf{v}_i^{\mathsf{cv}}\right|^2\right\}.
\end{align}
where we use the fact that $N_{\mathsf{AN}}=1$. By using a similar approach to \textit{Theorem 3}, we can also get
\begin{align}\label{eq:theo4:p4}
    &\sum_{i=1}^{N_{\mathsf{AN}}}\mathbb{E}\left\{|\textbf{h}_\mathsf{e}^{\rm H}\textbf{v}_i^{\mathsf{pp}}|^2\right\}\nonumber\\
    &=\frac{ \kappa }{ \kappa + 1 }\mathbb{E}\left\{\left|\textbf{h}_{\mathsf{LoS},\rm e}^{\rm H}\textbf{v}_i^{\mathsf{pp}}\right|^2\right\}+\frac{ 1 }{ \kappa + 1 }\mathbb{E}\left\{\left|\textbf{h}_{\mathsf{NLoS},\rm e}^{\rm H}\textbf{v}_i^{\mathsf{pp}}\right|^2\right\}\nonumber\\
    &\overset{(a)}{=}\frac{ \kappa }{ \kappa + 1 }\frac{1}{\mathbb{E}\left\{[(\mathbf{G}^{\rm H}\mathbf{G})^{-1}]_{\mathsf{ee}}\right\}}+\frac{ 1 }{ \kappa + 1 }\nonumber\\
    &\overset{(b)}{=}\frac{ \kappa }{ \kappa + 1 }\frac{N_{\mathsf{t}}-K-1}{[\boldsymbol{\Sigma}^{-1}_{\textbf{G}}]_{\mathsf{ee}}}+\frac{ 1 }{ \kappa + 1 },
\end{align}
where (a) comes from \eqref{eq:theo2:p1} and the subscription `$\mathsf{ee}$' indicates $(K+1)_{\mathsf{th}}$ diagonal entry of the matrix, and (b) comes from \eqref{eq:theo2:p7}. Note that $\textbf{v}_i^{\mathsf{pp}}$ is the $(K+1)_{\mathsf{th}}$ column vector of the ZF precoder of $\textbf{G}$ channel matrix and it correlates with the directional vector of UAV-Eve (see \eqref{eq:AN_precoder:prop}). By plugging \eqref{eq:theo4:p2}, \eqref{eq:theo4:p4} in \eqref{eq:rate_eve}, we can obtain \eqref{eq:theo4:def}. 

\bibliographystyle{IEEEtran} 
\bibliography{IEEEabrv,bibfile}
  
\end{document}